\documentclass[aps,pra,10pt,reprint]{revtex4-1}
\usepackage{amsmath} 
\usepackage{amsfonts}
\usepackage{amssymb}
\usepackage{bm} 
\usepackage{graphicx}
\usepackage[caption=false]{subfig}
\usepackage[colorlinks]{hyperref}

\DeclareMathOperator{\sgn}{sgn}

\begin{document}
\title{Spin wave radiation from vortices in \texorpdfstring{$^3$He}{3He}-B}
\author{S. M. Laine}
\email{sami.laine@oulu.fi}
\author{E. V. Thuneberg}
\affiliation{Nano and molecular systems research unit, University of Oulu, FI-90014, Finland}

\date{\today}

\begin{abstract}
We consider a vortex line in the B phase of superfluid $^3$He under uniformly precessing magnetization. The magnetization exerts torque on the vortex, causing its order parameter to oscillate. These oscillations generate spin waves, which is analogous to an oscillating charge generating electromagnetic radiation. The spin waves carry energy, causing dissipation in the system. Solving the equations of spin dynamics, we calculate the energy dissipation caused by spin wave radiation for arbitrary tipping angles of the magnetization and directions of the magnetic field, and for both vortex types of $^3$He-B. For the double-core vortex we also consider the anisotropy of the radiation and the dependence of the dissipation on twisting of the half cores. 
The radiated energy is compared with experiments in the mid-temperature range $T \sim 0.5 T_c$. The dependence of the calculated dissipation on several parameters is in good agreement with the experiments. Combined with numerically calculated vortex structure, the radiation theory produces the order of magnitude of the experimental dissipation. The agreement with the experiments indicates that spin wave radiation is the dominant dissipation mechanism for vortices in superfluid $^3$He-B in the mid-temperature range.
\end{abstract}

\maketitle

\section{Introduction}\label{sec:introduction}

Superfluid $^3$He is a useful paradigm of an unconventional superfluid or superconductor as it has a spin-triplet, p-wave-pairing order parameter which is precisely known. The B phase of superfluid $^3$He has  two well known vortex structures \cite{Ikkala82}. The {\em A-phase-core vortex} has A-phase-like order parameter in the vortex core \cite{Salomaa85}, while the {\em double-core vortex} has broken axisymmetry so that the vortex core is split into two half cores \cite{thuneberg1987}. Major part of the information about the superfluid phases and vortices in superfluid $^3$He has been obtained by nuclear magnetic resonance (NMR). The methods used are linear NMR with small tipping of the magnetization, as well as measurements using large tipping angles. The information about the order parameter is obtained by measuring either the frequency shift of resonance absorption or the amount of absorption, i.e., relaxation.

The purpose of our research is to understand the relaxation seen in NMR experiments on vortices of $^3$He-B. In particular, we consider experiments reported by Kondo et al.\ \cite{kondo1991,dmitriev1990}, which are made at intermediate temperatures around $0.5 T_c$ and at large tipping angles. A well known relaxation mechanism was first discussed by Leggett and Takagi \cite{Leggett77T}. It arises because the dipole-dipole interaction is enhanced by superfluid coherence but affects only the superfluid component, so that the normal and superfluid components of magnetization are driven out of mutual equilibrium. The conversion between the two components then leads to dissipation. This mechanism seems to be most effective at temperatures close to the transition temperature $T_c$. We find \cite{laine2016} that the Leggett-Takagi relaxation is too weak to explain the relaxation observed by Kondo et al. Another well known relaxation mechanism arises from diffusion of the normal component of the magnetization. It seems, however, that its contribution has to be small in the experiments by Kondo et al.\ since the observed magnetic field dependence \cite{dmitriev1990} is opposite to the one expected for spin diffusion. 

A third mechanism to cause relaxation was discussed in Ref.\ \cite{kondo1991}. It was suggested that the precessing magnetization drags the half cores of the double-core vortex to rotate around themselves. A phenomenological model for the rotation was constructed, and its parameters were fitted to the observed dissipation. More recently, the parameters of the rotational model were calculated  based on numerical solution of the vortex structure \cite{silaev2015}.  It was found that, while the rotation of the half cores was confirmed, the friction coefficient for the rotation is so large that the dissipation is negligible. Thus it remained open what causes the major part of the dissipation in the experiments by Kondo et al.\ \cite{kondo1991}.

In this paper we investigate a fourth relaxation mechanism. The precession of the magnetization makes the order parameter near the vortex to oscillate. These oscillations generate waves in the spin angular momentum, that is, {\em spin waves}. The spin waves radiated by the vortex carry energy and thus lead to relaxation of magnetization.

Generally, spin waves are collective excitations in systems possessing magnetic order. First predicted by Felix Bloch almost ninety years ago, they have recently become a subject of intense research in the fields of spintronics and magnonics because of their possible uses, e.g., in data transport and processing \cite{chumak2016}.
In superfluid $^3$He, the spin waves  were first detected by Osheroff et al.\ \cite{osheroff1977}, who saw standing spin wave modes in $^3$He-B. The first observations of vortices were based on the frequency shifts of such standing spin wave modes \cite{Ikkala82,Hakonen89}. The radiation of spin waves as a relaxation mechanism was discussed by Ohmi et al. \cite{Ohmi87} in connection with experiments in Ref.\ \cite{Ishikawa89}. The relaxation seen in the Josephson junction arrays of $^3$He has been interpreted in terms of spin wave radiation \cite{Viljas04}. The relaxation of magnetization by direct and parametric generation of spin waves has been reported by Zavjalov et al.\ \cite{Zavjalov2016}. Spin wave radiation from vortices has been pointed out by Volovik \cite{Volovik}. 

We study spin wave radiation from a single vortex under precessing magnetization. We consider arbitrary tipping  of magnetization, where we need to discuss separately tipping angles smaller and larger than the Leggett angle $\theta_L = \arccos(-1/4) \approx 104^\circ$. Besides a straight vortex, we also consider a vortex that is twisted by the precessing magnetization. By integrating the energy flux tensor around the vortex we calculate the total radiated energy. We find that radiation of spin waves is the dominant relaxation mechanism for vortices at low temperatures. The relaxation seen in the experiments by Kondo et al.\ \cite{kondo1991} can be well explained in terms of spin wave radiation. 

The paper is organized as follows.
In Sec.\ \ref{sec:static_vortex} we discuss the order parameter structure of static B-phase vortices far from the vortex core. In Sec.\ \ref{sec:spin_dynamics} we derive the equation of motion for the order parameter. The solution of this equation is discussed in Sec.\ \ref{sec:spin_waves}. In Sec.\ \ref{sec:energy} we calculate the energy carried by the spin waves. The effect of twisting of the vortex core on energy transport is studied in Sec.\ \ref{sec:twisted}. Finally, we compare the theory with experiments in Sec.\ \ref{sec:experiments}.

\section{Static vortex}\label{sec:static_vortex}

The order parameter of an isolated B-phase vortex far from the vortex axis can be written as  \cite{thuneberg1987,fogelstrom1995}
\begin{equation}\label{eq:orderParameterStatic}
\mathsf{A} = e^{i \varphi} \Delta_0 \mathsf{R} \left( \theta_0 \bm{\hat{n}}  \right) \mathsf{R} \left( \bm \theta  \right).
\end{equation}
Here $\Delta_0$ is the bulk gap and $\mathsf{R} \left( \theta_0 \bm{\hat{n}}  \right)$ is a finite rotation by an angle $\theta_0$ about an axis $\bm{\hat{n}}$. These, determined by the bulk, are assumed spatially constants. In the static case $\theta_0$ is fixed at the Leggett angle, $\theta_0 = \theta_L$.
The vortex appears through the phase $\varphi$, which equals the azimuthal angle around the vortex axis, and through an additional rotation $\mathsf{R} \left( \bm \theta  \right)$ by an angle $ \theta = \left| \bm \theta  \right|$ about an axis $ \bm{\hat{\theta}} = \bm \theta / \theta$.
The rotation $\bm \theta$ is determined by minimizing the free energy \cite{Leggett75,vollhardt1990,thuneberg2001}
\begin{equation}
F = \int_V dV \left( f_D + f_G  \right)
\end{equation}
in the region $V$ excluding the vortex core with appropriate boundary conditions.
Here $f_D$ originates from the dipole-dipole interaction between the $^3$He nuclei,
\begin{equation}\label{eq:f_D_static}
f_D = \lambda_D \left(  R_{i i} R_{j j} + R_{i j} R_{j i}  \right) \approx - \frac{\lambda_D}{2} + \frac{15}{2} \lambda_D \left( \bm{\hat{n}} \cdot \bm \theta  \right)^2,
\end{equation} 
while $f_G$ is the gradient energy,
\begin{equation}\label{eq:f_G_static}
\begin{split}
f_G &= \lambda_{G1} \frac{\partial R_{\alpha i}}{\partial r_i} \frac{\partial R_{\alpha j}}{\partial r_j} + \lambda_{G2} \frac{\partial R_{\alpha j}}{\partial r_i} \frac{\partial R_{\alpha j}}{\partial r_i} \\
&\approx 2 \lambda_{G2} \left[ (1 + c) \partial_i \theta_k  \partial_i \theta_k - c \partial_i \theta_k  \partial_k \theta_i \right].
\end{split}
\end{equation}
In the above we have denoted $\mathsf{R} = \mathsf{R} \left( \theta_0 \bm{\hat{n}}  \right) \mathsf{R} \left( \bm \theta  \right)$ and $c = \lambda_{G1} / 2 \lambda_{G2}$. The coefficients $\lambda_D$, $\lambda_{G1}$, and $\lambda_{G2}$ depend on temperature and pressure.
We assume $\bm \theta$ to be small so that the energies are well approximated by expressions that are quadratic in $\bm \theta$.
The gradient energy dominates the dipole energy when the distance from the vortex core is  much less than the dipole length $\xi_D = \sqrt{\lambda_{G2} / \lambda_D}$.
If we neglect the dipole energy, the solution describing an isolated vortex is $\bm \theta = \bm \theta_v$, where \cite{silaev2015, laine2016}
\begin{equation}\label{eq:theta_v}
\begin{split}
\bm \theta_v (r, \varphi) &= \frac{C_1 \cos \varphi}{r} \left( \frac{\sin \varphi}{1+c} \bm{\hat{r}} + \cos \varphi \bm{\hat{\varphi}}  \right) \\
&- \frac{C_2 \sin \varphi}{r} \left( \frac{\cos \varphi}{1+c} \bm{\hat{r}} - \sin \varphi \bm{\hat{\varphi}}  \right).
\end{split}
\end{equation}
Here $\bm{\hat{r}}$, $\bm{\hat{\varphi}}$, and $\bm{\hat{z}}$ are the basis vectors of cylindrical coordinate system with $\bm{\hat{z}}$ oriented along the vortex axis. This is a good approximation at distances $10\, \xi(T) \lesssim r \ll \xi_D$ from the axis, where $\xi(T)$ is the temperature dependent coherence length. Near the core $\theta$ becomes large and the second order expansion of the gradient energy breaks down. The inclusion of the dipole energy causes $\bm \theta$ to vanish more rapidly than $r^{-1}$ at distances greater than $\xi_D$. 

The coefficients $C_1$ and $C_2$ depend on the type of the vortex. They can be extracted from the numerical solution of the vortex core structure \cite{thuneberg1987,fogelstrom1995,silaev2015}. Because of axial symmetry, $C_1 = C_2$ for the A-phase-core vortex and thus $\bm \theta_v = C_1 \bm{\hat{\phi}} / r$. This special case of Eq.\ (\ref{eq:theta_v}) was found by Hasegawa \cite{hasegawa1985}. For the double-core vortex $C_1 / C_2 \gg 1$. 

\begin{figure}[tb]
\centering
\subfloat[]{\label{fig:staticTheta_A} \includegraphics[width=0.49\columnwidth]{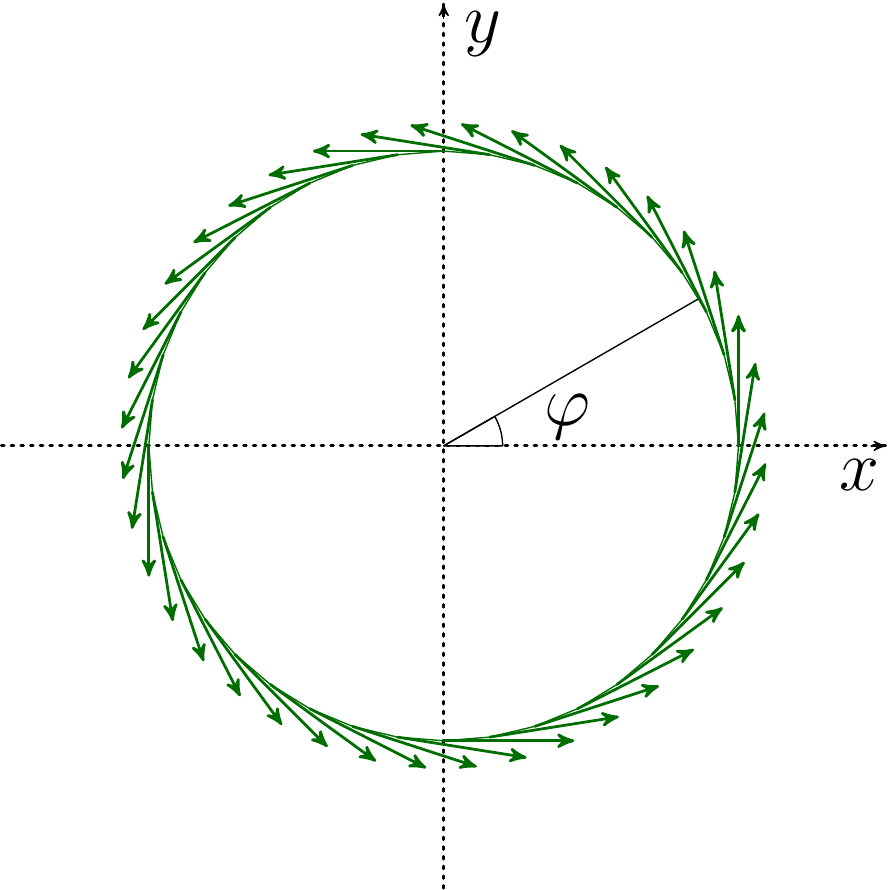}}
\subfloat[]{\label{fig:staticTheta_D} \includegraphics[width=0.49\columnwidth]{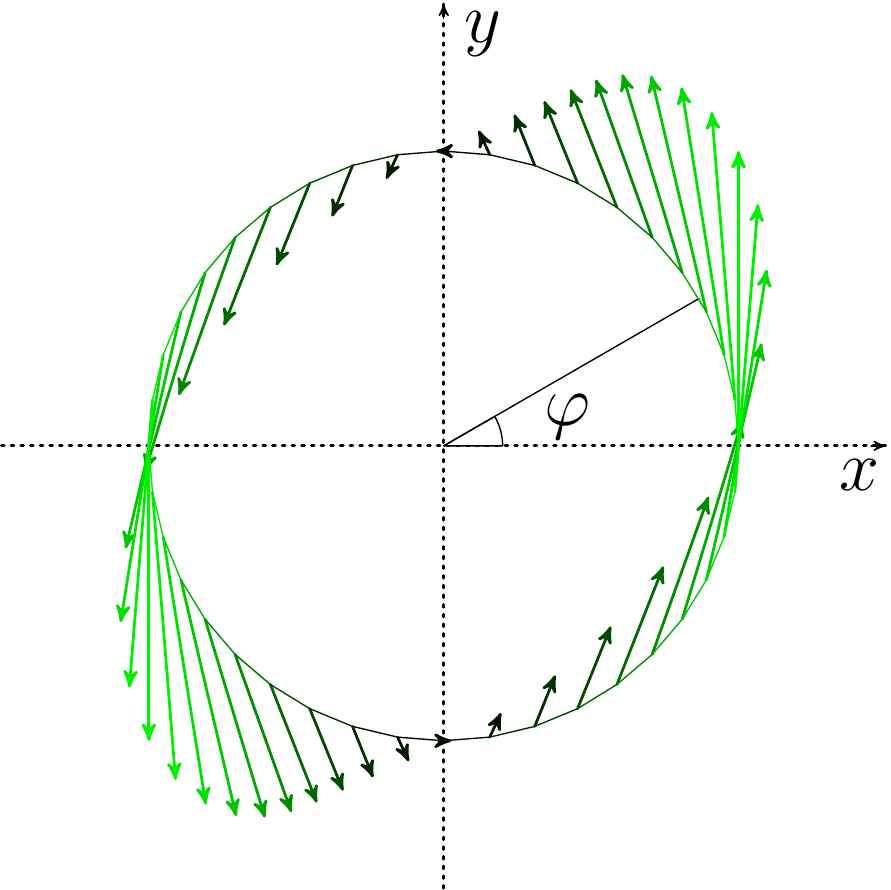}} 
\caption{The structure of $r \bm \theta_v(r, \varphi) \equiv \bm \vartheta_v(\varphi)$ plotted on a circle for (a) the A-phase-core vortex and (b) the double-core vortex. The unit of distance is arbitrary but equal in both (a) and (b). The parameters used correspond to $T = 0.6 T_c$, $p = 29.3$ bar, and $c = 1$.}
\label{fig:staticTheta}
\end{figure}

Since $\bm \theta_v \propto r^{-1}$ to leading order, $r \bm \theta_v$ is independent of $r$. Thus we can visualize $\bm \theta_v$ by plotting $r \bm \theta_v(r, \varphi) \equiv \bm \vartheta_v(\varphi)$ on a circle in the $xy$-plane. This is shown in Fig.\ \ref{fig:staticTheta}. We have used the values $C_1 = C_2 = 1.33 R_0$ for the A-phase-core vortex and $C_1 = 3.00 R_0$, $C_2 = 0.08 R_0$ for the double-core vortex. Here $R_0 = (1 + F_1^s / 3) \xi_0$, $F_1^s$ is a Fermi liquid parameter, $\xi_0 = \hbar v_F / 2 \pi k_B T_c$ is the coherence length, $v_F$ is the Fermi velocity, and $T_c$ is the critical temperature. The values of $C_1$ and $C_2$ correspond to temperature $T = 0.6 T_c$ and pressure $p = 29.3$ bar \cite{silaev2015}. We have also set $c = 1$ since this is the weak-coupling value assuming vanishing Fermi-liquid parameters $F_1^a$ and $F_3^a$.
The structure of the A-phase-core vortex is shown in Fig.\ \ref{fig:staticTheta}(a).
The structure of the double-core vortex is shown in Fig.\ \ref{fig:staticTheta}(b). The half cores are located on the $y$-axis.

\section{Spin dynamics}\label{sec:spin_dynamics}

We now place the vortex in a static external magnetic field $\bm B$ and study spin dynamics. This is governed by the Leggett theory \cite{leggett1974}.
Within the Leggett theory, the motion of the order parameter in spin space is purely rotational, $\mathsf{A}(t) = \mathsf{R}(t) \mathsf{A}_0$. Here $\mathsf{R}$ is a time-dependent rotation matrix and $\mathsf{A}_0$ is the initial order parameter. Since the order parameter of the vortex is of the correct form, see Eq.\ (\ref{eq:orderParameterStatic}), we can include the time-dependence of $\mathsf{A}$ in variables $\theta_0$, $\bm{\hat{n}}$ and $\bm \theta$. This means that the dynamic order parameter is of the same form as the static one,
\begin{equation}\label{eq:orderParameterDynamic}
\mathsf{A} = e^{i \varphi} \Delta_0 \mathsf{R} \left( \theta_0 \bm{\hat{n}}  \right) \mathsf{R} \left(  \bm \theta  \right) .
\end{equation}
We study a holonomically constrained problem (see, e.g., \cite{fetter1980}) where $\theta_0(t)$ and $\bm{\hat{n}}(t)$ are given functions of time. They are determined by the bulk, i.e.,\ they solve the equations of spin dynamics in the absence of the vortex. We take the bulk solution to be the Brinkman-Smith (BS) mode \cite{Brinkman75b,Fomin83}, where the magnetization precesses uniformly about $\bm B$. Details of the Brinkman-Smith mode are discussed later. The {\em system} we want to study is the vortex, described by the field $\bm \theta(\bm r, t)$. The Brinkman-Smith mode then acts as an external {\em drive} for the system. In order to maintain the Brinkman-Smith mode in the presence of the vortex, energy is needed from an outside source. Experimentally this is done using a time-dependent magnetic field. In our calculations the energy source is present implicitly through the constraints.

\begin{figure}[tb]
\centering
\includegraphics[width=0.48\textwidth]{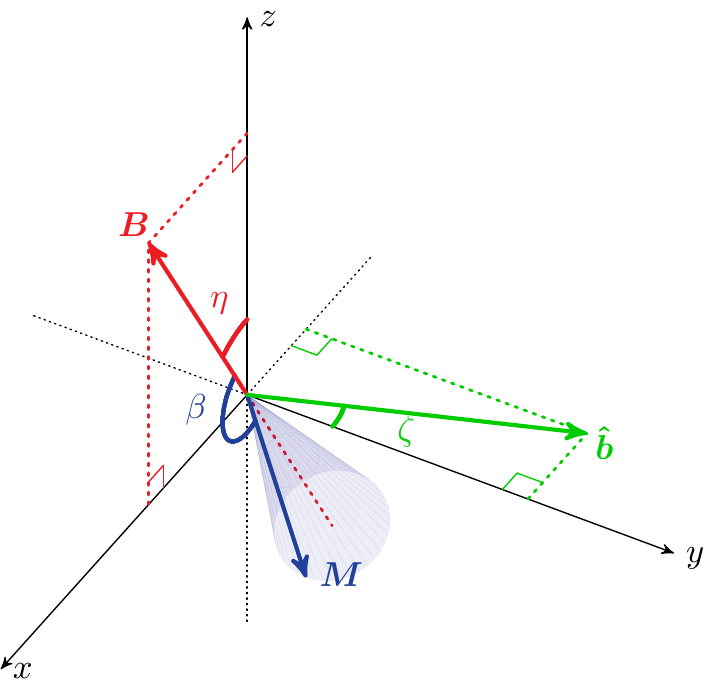}
\caption{Definitions of the angles $\beta$, $\zeta$, and $\eta$. The vortex axis coincides with the $z$-axis. $\bm B$ is the external static magnetic field, $\bm{\hat{b}}$ is the anisotropy vector of the double-core vortex, pointing from one of the half cores to the other, and $\bm M$ is the magnetization density. Note that $\bm M$ is not static but precesses uniformly about $\bm B$ with tipping angle $\beta$.}
\label{fig:geometry}
\end{figure}

We shall use the following geometry and notation. The $z$-axis of the coordinate system coincides with the vortex axis.
The $x$-axis is chosen so that the magnetic field $\bm B$ lies in the $xz$-plane. In addition to the cartesian coordinate system $(x,y,z)$ we shall use the standard cylindrical coordinate system $(r, \varphi, z)$, where $r$ is the distance from the $z$-axis and $\varphi$ is the azimuthal angle, measured anticlockwise from the $x$-axis. The tilting angle of $\bm B$ from the vortex axis is denoted by $\eta$, so that $\bm{\hat{B}} = \cos \eta \bm{\hat{z}}  + \sin \eta \bm{\hat{x}} $. The orientation of the vortex core in the $xy$-plane is described by an angle $\zeta$. More specifically, the anisotropy vector $\bm{\hat{b}}$ of the double-core vortex, pointing from one of the half cores to the other, is given by $\bm{\hat{b}} = \cos \zeta \bm{\hat{y}} - \sin \zeta \bm{\hat{x}}$. Since the A-phase-core vortex is cylindrically symmetric, there is no need to define its orientation. Finally, $\beta$ is the tipping angle of the magnetisation $\bm M$, measured from the direction of the magnetic field, so that $\cos \beta = \bm{\hat{M}} \cdot \bm{\hat{B}}$. Figure \ref{fig:geometry} shows the definitions of the various quantities in graphical form.

As mentioned above, in Brinkman-Smith mode the magnetization precesses uniformly about $\bm B$,
\begin{equation}
\bm M_{BS} = M_{BS} \mathsf{R}(\eta \bm{\hat{y}}) \mathsf{R}(\omega_{BS} t \bm{\hat{z}}) \mathsf{R}(\beta \bm{\hat{y}}) \cdot \bm{\hat{z}}.
\end{equation}
The precession rate is given by
\begin{equation}\label{eq:BrinkmanSmithOmega}
\omega_{BS} = \frac{\omega_L}{2} \left( 1 + \sqrt{ 1 - \frac{16 \Omega^2}{15 \omega_L^2} \left( 1 + 4 \cos \theta_0  \right)} \right) .
\end{equation}
Here $\omega_L = - \gamma_0 B$ is the Larmor frequency, $\Omega$ is the longitudinal NMR frequency, $\Omega^2 = 15 \mu_0 \gamma_0^2 \lambda_{D} / \chi$, $\mu_0$ is the vacuum permeability, $\gamma_0$ is the gyromagnetic ratio of $^3$He, and $\chi$ is the magnetic susceptibility of the B phase. The rotation angle $\theta_0$ is independent of time.
If the magnetization of the sample is tipped by an angle $\beta \leq \theta_L$, then $\theta_0 = \theta_L$ and $\omega_{BS} = \omega_L$. If $\beta > \theta_L$, then $\theta_0$ satisfies
\begin{equation}
\cos \beta = \frac{ \omega_{BS} ( \cos \theta_0 - 1) / \omega_L + 1 }{ \sqrt{\omega_{BS}^2 \sin^2 \theta_0 / \omega_L^2 + \left[ \omega_{BS} ( \cos \theta_0 - 1) / \omega_L + 1 \right]^2 }}.
\end{equation}
This means that $\theta_0 > \theta_L$, and so the precession rate is increased, $\omega_{BS} > \omega_L$.
The unit vector $\bm{\hat{n}}$ precesses uniformly about $\bm B$ with the same rate as the magnetization. It can be written as
\begin{equation}\label{eq:nt}
\bm{\hat{n}}(t) = \mathsf{R}(\eta \bm{\hat{y}}) \mathsf{R}(\omega_{BS} t \bm{\hat{z}}) \cdot \bm{\hat{n}}_0,
\end{equation}
where
\begin{equation}
\begin{split}
\bm{\hat{n}}_0 &= 
\begin{cases}
\frac{2}{\sqrt{5}}\sqrt{1 - \cos \beta} \bm{\hat{y}} + \frac{1}{\sqrt{5}}\sqrt{1 + 4 \cos \beta} \bm{\hat{z}}, &  \beta \leq \theta_L \\
\bm{\hat{y}}, & \beta > \theta_L
\end{cases}.
\end{split}
\end{equation}

There are different ways to proceed, but a convenient one in our case is the Lagrangian formulation \cite{maki1975,maki1977}. 
As in mechanics, there is an angular velocity $\bm \omega$ related to the rotating motion of the order parameter, defined by
\begin{equation}\label{eq:omega_def}
\dot{R}_{\alpha i} = \varepsilon_{\alpha \beta \gamma} \omega_{\beta} R_{\gamma i}.
\end{equation}
Here a dot over a letter denotes differentiation with respect to time, $ \varepsilon_{\alpha \beta \gamma}$ is the Levi-Civita symbol, and $\mathsf{R} = \mathsf{R} \left( \theta_0 \bm{\hat{n}}  \right) \mathsf{R} \left( \bm \theta  \right)$.
In terms of $\bm \omega$, the Lagrangian density of the system can be written as
\begin{equation}\label{eq:lagrangian}
\mathcal{L}  = \frac{1}{2 \mu_0 \gamma_0^2} \left( \bm \omega - \bm \omega_L \right) \cdot \bm{\stackrel{\leftrightarrow}{\chi}} \cdot \left( \bm \omega - \bm \omega_L \right) - f_D - f_G,
\end{equation}
where $\bm{\stackrel{\leftrightarrow}{\chi}}$ is the magnetic susceptibility tensor and $\bm \omega_L = - \gamma_0 \bm B = \omega_L \bm{\hat{B}}$ is the Larmor frequency vector.
In addition to the angular velocity, one can also define the generalised momentum canonically conjugate to the rotation. This is the spin density $\bm S = \bm M / \gamma_0$. The vectors $\bm \omega$ and $\bm S$ are related by
\begin{equation}\label{eq:spin_density}
\bm S = \frac{\partial \mathcal{L}}{\partial \bm \omega} = \frac{\bm{\stackrel{\leftrightarrow}{\chi}}}{\mu_0 \gamma_0^2} \cdot \left( \bm \omega - \bm \omega_L \right).
\end{equation}
It follows from Eqs.\ (\ref{eq:nt}) and (\ref{eq:omega_def}) that
\begin{equation}\label{eq:omega_second_order}
\bm \omega \approx \bm \omega_{BS} + \mathsf{R}\left( \theta_0 \bm{\hat{n}} \right) \cdot \left( \bm{\dot{\theta}} + \frac{1}{2} \bm \theta \times \bm{\dot{\theta}}  - \bm \omega_{BS}   \right),
\end{equation}
where $\bm \omega_{BS} = \omega_{BS} \bm{\hat{B}}$.
Here we have kept again only the two lowest order terms in $\bm \theta$. 
Since $\theta_0$ is not necessarily equal to $\theta_L$, the second order expansion of the dipole energy is modified from Eq.\ (\ref{eq:f_D_static}) to
\begin{equation}
\begin{split}
f_D / \lambda_D &\approx 4 \cos \theta_0 (1 + 2 \cos \theta_0) \\
&- 4 \sin \theta_0 (1 + 4 \cos \theta_0)(\bm{\hat{n}} \cdot \bm \theta) \\
&- (1 + \cos \theta_0 )(1 + 4 \cos \theta_0) (\bm \theta \cdot \bm \theta) \\
&+ 3 (1 - \cos \theta_0)(4 \cos \theta_0 + 3)(\bm{\hat{n}} \cdot \bm \theta)^2.
\end{split}
\end{equation}
The expansion of the gradient energy is still given by Eq.\ (\ref{eq:f_G_static}).
Substituting these into the Lagrangian density and using the fact that the susceptibility in the B phase is diagonal, $\chi_{\mu \nu} = \chi \delta_{\mu \nu}$,
we derive the linearized equation of motion for $\bm \theta$ using the formula familiar from classical field theory \cite{fetter1980},
\begin{equation}
\frac{\partial}{\partial t} \frac{\partial \mathcal{L}}{\partial \dot{\theta_i}} + \partial_j  \frac{\partial \mathcal{L}}{\partial \partial_j \theta_i} - \frac{\partial \mathcal{L}}{\partial \theta_i} = 0.
\end{equation}
As a result we get
\begin{equation}\label{eq:EquationOfMotion}
\begin{split}
\bm{\ddot{\theta}} &- \omega_{BS} \bm w \times \bm{\dot{\theta}} + \Omega^2 \mathsf{L} \cdot \bm \theta \\
&- v^2 \left[(1+c) \nabla^2 \bm{\theta} - c \bm \nabla \left( \bm \nabla \cdot \bm{\theta} \right) \right] = 0.
\end{split}
\end{equation}
Here $v$ is a characteristic spin wave velocity, defined by $v^2 = 4 \mu_0 \gamma_0^2 \lambda_{G2} / \chi $,
\begin{equation}
\bm w = \bm{\hat{B}} - \frac{\omega_{BS} - \omega_L}{\omega_{BS}} \mathsf{R}^T(\theta_0 \bm{\hat{n}}) \cdot \bm{\hat{B}},
\end{equation}
and
\begin{equation}
\begin{split}
\mathsf{L} \cdot \bm \theta &= \frac{2}{15} \sin \theta_0 (1 + 4 \cos \theta_0) \bm{\hat{n}} \times \bm \theta  \\
&- \frac{2}{15} (1 + \cos \theta_0 )(1 + 4 \cos \theta_0) \bm \theta \\
&+ \frac{2}{5} (1 - \cos \theta_0)(4 \cos \theta_0 + 3) \bm{\hat{n}} (\bm{\hat{n}} \cdot \bm \theta).
\end{split}
\end{equation}

In the rest of the paper we shall work with dimensionless quantities, unless stated otherwise. We take the unit of length to be $v / \Omega = 2 \xi_D / \sqrt{15}$ and the unit of time to be $\Omega^{-1}$. We measure the angular frequencies $\omega_L$ and $\omega_{BS}$ in units of $\Omega$ and the coefficients $C_1$ and $C_2$ in units of $R_0$. Finally, we measure $\bm \theta$ in (dimensionless) units of $R_0 \Omega / v$.
In these units Eq.\ (\ref{eq:EquationOfMotion}) can be written as
\begin{equation}\label{eq:EquationOfMotionDimensionless}
\begin{split}
&\bm{\ddot{\theta}}  - \omega_{BS} \bm w \times \bm{\dot{\theta}} + \mathsf{L} \cdot \bm \theta - (1+c) \nabla^2 \bm{\theta} + c \bm \nabla \left( \bm \nabla \cdot \bm{\theta} \right) = 0.
\end{split}
\end{equation}

\section{Spin waves}\label{sec:spin_waves}

In this section we solve the equation of motion (\ref{eq:EquationOfMotionDimensionless}) in two different approximations. We shall see that in both cases the solution contains a part representing waves propagating away from the vortex. Since $\bm \theta$ and $\bm S$ are coupled via Eqs.\ (\ref{eq:spin_density}) and (\ref{eq:omega_second_order}), this means that the vortex radiates spin waves.
Physically this stems from two factors.
First, due to the dipole interaction, the rotating $\bm{\hat{n}}$ exerts torque on $\bm \theta$ at each point in space, causing it to oscillate with time. 
Second, because of the gradient energy, $\bm \theta$ at each point is strongly coupled to its neighbouring points. This means that any disturbances in $\bm \theta$ are propagated in space.

To obtain a solution which properly describes a vortex, we split $\bm \theta$ into two parts,
\begin{equation}
\bm \theta (\bm r, t) = \bm \theta_1(\bm r) + \bm \theta_2(\bm r, t).
\end{equation}
Here $\bm \theta_1(\bm r)$ is a static solution with correct behaviour near the core and $\bm \theta_2(\bm r, t)$ is a time-dependent deviation from the static solution. To ensure that the solution has the correct form near the core, we demand that $\bm \theta_2$ vanishes when $r \to 0$.
From Eq.\ (\ref{eq:EquationOfMotionDimensionless}) we then obtain
\begin{equation}\label{eq:EquationOfMotionSource}
\begin{split}
\bm{\ddot{\theta}}_2  &- \omega_{BS} \bm w \times \bm{\dot{\theta}}_2 + \mathsf{L} \cdot \bm \theta_2 \\
 &- (1+c) \nabla^2 \bm{\theta}_2 + c \bm \nabla \left( \bm \nabla \cdot \bm{\theta}_2 \right)= \bm \rho \left( \bm r, t \right) ,
\end{split}
\end{equation}
where
\begin{equation}\label{eq:rho}
\bm \rho \left( \bm r, t \right) = -\mathsf{L} \cdot \bm \theta_1 +  (1+c) \nabla^2 \bm{\theta}_1 - c \bm \nabla \left( \bm \nabla \cdot \bm{\theta}_1 \right).
\end{equation}

Since the gradient energy dominates the dipole energy near the vortex core, we take $\bm \theta_1$ to minimize the gradient free energy. Taking into account the orientation of the vortex we have
\begin{equation}\label{eq:theta_1}
\bm \theta_1\left( r, \varphi  \right) = \mathsf{R} ( \zeta \bm{\hat{z}}  ) \cdot \bm \theta_v (r, \varphi - \zeta),
\end{equation}
where $\bm \theta_v (r, \varphi)$ is given by Eq.\ (\ref{eq:theta_v}). The source term simplifies to $\bm \rho \left( \bm r, t \right) = -\mathsf{L}(t) \cdot \bm \theta_1(\bm r)$.
In Sec.\ \ref{sec:twisted} we study the effect of twisting of the vortex core. There we still use Eq.\ (\ref{eq:theta_1}), but with $\zeta = \zeta(z)$. This means that we have to keep the full expression (\ref{eq:rho}).

Written componentwise, Eq. (\ref{eq:EquationOfMotionSource}) is a system of three coupled second-order inhomogeneous linear partial differential equations. Solving this is not trivial because of the time-dependent coefficients $\mathsf{L}(t)$ and $\bm w(t)$ and the non-laplacian gradient term $\propto c\bm\nabla\bm\nabla$. While the time dependence of $\mathsf{L}$ and $\bm w$ can be removed by transformation to a frame rotating in the spin space, the non-laplacian operator is complicated there because it is anisotropic in separate spin or orbit space rotations. In the following we study two alternative approximations. In case A we set $c = 0$. This removes the non-laplacian term and allows solution in the rotating frame. In case B we consider the limit of high magnetic field, $\omega_L \gg 1$. In this limit we may neglect the time-dependence of the coefficients $\mathsf{L}(t)$ and $\bm w(t)$ on the left-hand side of Eq.\ (\ref{eq:EquationOfMotionSource}).

When solving the equation of motion, we shall work partly in two dimensional Fourier space. We use the convention
\begin{align}
f(\bm k) &= \iint d^2 r \exp \left( - i \bm k \cdot \bm r \right) f(\bm r), \\
f(\bm r) &= \frac{1}{(2 \pi)^2} \iint d^2 k \exp \left( i \bm k \cdot \bm r \right) f(\bm k),
\end{align}
with $\bm r = x \bm{\hat{x}} + y \bm{\hat{y}} = r \cos \varphi \bm{\hat{x}} + r \sin \varphi \bm{\hat{y}}$ and $\bm k = k_x \bm{\hat{x}} + k_y \bm{\hat{y}} = k \cos \varphi_k \bm{\hat{x}} + k \sin \varphi_k \bm{\hat{y}}$.
The equation of motion in the Fourier space is then obtained by making a substitution $\bm \nabla \to i \bm k$. Using the definition above, the Fourier transform of $\bm \theta_v$, Eq.\ (\ref{eq:theta_v}), is given by
\begin{equation}\label{eq:theta_v_Fourier}
\begin{split}
\bm \theta_v (k, \varphi_k) = &-\frac{2 \pi i}{k} \frac{C_1 - C_2}{2} \sin(2 \varphi_k) \bm{\hat{k}} \\
& -\frac{2 \pi i}{k}  \left[ \frac{C_1 + C_2}{2} + \frac{C_1 - C_2}{2(1 + c)} \cos(2 \varphi_k)  \right] \bm{\hat{\varphi}}_k  .
\end{split}
\end{equation}
Here $\bm{\hat{k}} = \cos \varphi_k \bm{\hat{x}} + \sin \varphi_k \bm{\hat{y}}$ and $\bm{\hat{\varphi}}_k = -\sin \varphi_k \bm{\hat{x}} + \cos \varphi_k \bm{\hat{y}}$ are the basis vectors of polar coordinate system in Fourier space.

\subsection{Isotropic approximation}

We start by considering Eq.\ (\ref{eq:EquationOfMotionSource}) in the limit $c = 0$.
Since the vector $\bm{\hat{n}}$ rotates about $\bm B$ at constant rate, it is convenient to use a basis where $\bm{\hat{n}}$ is constant.
Mimicking the form of $\bm{\hat{n}}$ in Eq.\ (\ref{eq:nt}) we define
\begin{equation}
\bm \theta_2 \left( \bm r, t \right) = \mathsf{R} ( \eta \bm{\hat{y}} ) \mathsf{R} ( \omega_{BS} t \bm{\hat{z}} ) \cdot \bm \alpha \left( \bm r, t \right).
\end{equation}
The equation of motion for $\bm \alpha$ is then
\begin{equation}\label{eq:EquationOfMotionRotatingBasis}
\begin{split}
\bm{\ddot{\alpha}} &+ \omega_{BS} \bm{\hat{z}} \times \bm{\dot{\alpha}} - \bm w_0 \times \left( \bm{\dot{\alpha}} + \omega_{BS} \bm{\hat{z}} \times \bm \alpha  \right) \\
&+ \mathsf{L}_0 \cdot \bm{\alpha} - \nabla^2 \bm {\alpha} = - \mathsf{R} ( - \omega_{BS} t \bm{\hat{z}} ) \mathsf{R} ( - \eta \bm{\hat{y}} ) \cdot \mathsf{L} \cdot \bm \theta_1,
\end{split}
\end{equation}
where we have defined
\begin{equation}
\begin{split}
\mathsf{L}_0 \cdot \bm \alpha &= \frac{2}{15} \sin \theta_0 (1 + 4 \cos \theta_0) \bm{\hat{n}}_0 \times \bm \alpha  \\
&- \frac{2}{15} (1 + \cos \theta_0 )(1 + 4 \cos \theta_0) \bm \alpha \\
&+ \frac{2}{5} (1 - \cos \theta_0)(4 \cos \theta_0 + 3) \bm{\hat{n}}_0 (\bm{\hat{n}}_0 \cdot \bm \alpha)
\end{split}
\end{equation}
and
\begin{equation}
\bm w_0 = \frac{4}{15} \omega_{BS}^{-1} (1 + 4 \cos \theta_0) \mathsf{R}^T(\theta_0 \bm{\hat{n}}_0) \cdot \bm{\hat{z}}.
\end{equation}
Note that the coefficients on the left-hand side of (\ref{eq:EquationOfMotionRotatingBasis}) are independent of time. This happens only when $c = 0$.
The source term on the right-hand side of (\ref{eq:EquationOfMotionRotatingBasis}) can be written as
\begin{equation}
\begin{split}
-\mathsf{R} ( - \omega_{BS} t \bm{\hat{z}} ) \mathsf{R} ( - \eta \bm{\hat{y}} ) \cdot \mathsf{L} \cdot \bm \theta_1 = \Re \left\{ \bm \rho_0(\bm r) + e^{- i \omega_{BS} t} \bm \rho_1 (\bm r)   \right\},
\end{split}
\end{equation}
where
\begin{align}
\bm \rho_0(\bm r) &= -\mathsf{L}_0 \cdot \mathsf{M}_0 \cdot \mathsf{R} ( - \eta \bm{\hat{y}} ) \cdot \bm \theta_1 (\bm r), \\
\bm \rho_1(\bm r) &= -\mathsf{L}_0 \cdot \mathsf{M}_- \cdot \mathsf{R} ( - \eta \bm{\hat{y}} ) \cdot \bm \theta_1 (\bm r),
\end{align}
and
\begin{align}
&\mathsf{M}_0 = 
\begin{pmatrix}
0 & 0 & 0 \\
0 & 0 & 0 \\
0 & 0 & 1
\end{pmatrix},
&\mathsf{M}_- = 
\begin{pmatrix}
1 & i & 0 \\
-i & 1 & 0 \\
0 & 0 & 0
\end{pmatrix}.
\end{align}

We now make a complex ansatz
\begin{equation}
\bm \alpha (r, \varphi) = \bm \alpha_0 (r, \varphi) + e^{-i \omega_{BS} t} \bm \alpha_1 (r, \varphi),
\end{equation}
the real part of which is the physical solution, and obtain the equations
\begin{align}
\mathsf{K}_0(\bm  \nabla) \cdot \bm \alpha_0 (\bm r)  &= \bm \rho_0 (\bm r), \\
\mathsf{K}_1(\bm \nabla) \cdot \bm \alpha_1 (\bm r)  &= \bm \rho_1 (\bm r),
\end{align}
where
\begin{align}
\mathsf{K}_0(\bm \nabla) &= - \nabla^2 + \mathsf{L}_0 - \omega_{BS} [\bm w_0]_\times \cdot [\bm{\hat{z}}]_\times, \\
\mathsf{K}_1(\bm \nabla) &= \mathsf{K}_0(\bm \nabla) - \omega_{BS}^2 \mathsf{I} - i \omega_{BS} [\omega_{BS} \bm{\hat{z}} - \bm w_0]_\times.
\end{align}
Here $[\bm w]_\times \cdot \bm v \equiv \bm w \times \bm v$ and $\mathsf{I}$ is the identity operator.
Since these equations are linear, it is convenient to solve them first in the Fourier space and then transform back to the coordinate space. 
In Fourier space the two PDEs are transformed into algebraic equations which are easily solved for $\bm \alpha_0 (\bm k)$ and $\bm \alpha_1 (\bm k)$,
\begin{align}
\bm \alpha_0 (\bm k) &= \mathsf{K}^{-1}_0(i \bm k) \cdot \bm \rho_0(\bm k) = \theta_{1x}(\bm k) \sin \eta \frac{\bm D_0}{k^2 - k_0^2}, \label{eq:sol_alpha_kspace_0}\\
\bm \alpha_1 (\bm k) &= \mathsf{K}^{-1}_1(i \bm k) \cdot \bm \rho_1(\bm k) \nonumber \\
&= \left[  \theta_{1y}(\bm k) - i \theta_{1x}(\bm k) \cos \eta \right] \sum_{j = 1}^{3} \frac{\bm D_j}{k^2 - k_j^2} \label{eq:sol_alpha_kspace_1}.
\end{align}
Here $\bm D_m$ and $k_m^2$, $m = 0,1,2,3$, are obtained using partial fraction decomposition with respect to $k^2$.
It can be seen that $k_0^2$ is always negative, $k_2^2$ and $k_3^2$ are always positive, and $k_1^2$ is negative at $\beta \lesssim 140^\circ$, changing to positive at larger values of the tipping angle.
The explicit expressions of $\bm D_m$ and $k_m^2$ are too cumbersome to be written down here. 

Next we take the inverse transform of $\bm \alpha_i(\bm k)$,
\begin{equation}
\begin{split}
\bm \alpha_i(\bm r) &= \frac{1}{\left( 2 \pi \right)^2} \int_0^{2 \pi} d \varphi_k \int_0^\infty k dk e^{i k r \cos(\varphi - \varphi_k)} \bm \alpha_i(\bm k) \\
&= \frac{1}{\left( 2 \pi \right)^2} \int_0^{2 \pi} d \varphi_k \int_0^\infty k dk \Bigg\{ J_0(k r) \\
&\hphantom{=}+ 2 \sum_{n = 1}^\infty i^n J_n(k r) \cos \left[ n (\varphi - \varphi_k)  \right]  \Bigg\} \bm \alpha_i(\bm k).
\end{split}
\end{equation}
Here we have used the Jacobi-Anger expansion \cite{abramowitz1965} to expand the exponential.
Using $\bm \alpha_i(\bm k)$ from Eqs.\ (\ref{eq:sol_alpha_kspace_0}) and (\ref{eq:sol_alpha_kspace_1}), and $\bm \theta_1$ from Eq.\ (\ref{eq:theta_1}), we have
\begin{align}
\bm \alpha_0 (\bm r) &= \vartheta_{1x}(\varphi) \sin \eta \bm D_0 \int_0^\infty dk \frac{J_1(k r)}{k^2 - k_0^2}, \\
\bm \alpha_1 (\bm r) &= \left[  \vartheta_{1y}(\varphi) - i \vartheta_{1x}(\varphi) \cos \eta \right] \sum_{j = 1}^{3} \bm D_j \int_0^\infty dk  \frac{J_1(k r)}{k^2 - k_j^2},
\end{align}
where $\bm \vartheta_1(\varphi) \equiv r \bm \theta_1(r, \varphi)$.

The next step is to evaluate the integral $\int_0^\infty dk J_1(k r) / (k^2 - k_m^2)$. If $k_m^2<0$, the integrand is finite on the positive $k$-axis and can be evaluated analytically \cite{abramowitz1965}.
If $k_m^2>0$, there is a simple pole at $k = k_m$. In this case we use the standard trick and shift the pole slightly away from the real axis by adding a small imaginary part to the denominator, $k_m \to k_m \pm i \varepsilon$, $\varepsilon > 0$. The choice of sign here determines the asymptotic behaviour of the solution. We choose the positive sign since this makes the solution an outward travelling wave. The negative sign would lead to a wave travelling towards the vortex. After evaluating the integral we take the limit $\varepsilon \to 0$. As a result we get
\begin{align}
\bm \alpha_0 (\bm r) &= \vartheta_{1x}(\varphi) \sin \eta \frac{\bm D_0}{k_0} \left[ \frac{i \pi}{2} H_1^{(1)}(k_0 r)  -\frac{1}{k_0 r} \right], \\
\bm \alpha_1 (\bm r) &= \left[ \vartheta_{1y}(\varphi) - i \vartheta_{1x}(\varphi) \cos \eta \right] \nonumber \\
&\times \sum_{j = 1}^{3} \frac{\bm D_j}{k_j} \left[ \frac{i \pi}{2} H_1^{(1)}(k_j r)  -\frac{1}{k_j r} \right].
\end{align}
Here $H_1^{(1)}(x)$ is a Hankel function of the first kind.
Note that both $\bm \alpha_0 (\bm r)$ and $\bm \alpha_1 (\bm r)$ are zero at the origin. This means that the behaviour of $\bm \theta (\bm r)$ near the core is determined by $\bm \theta_1 (\bm r)$, as was claimed earlier.
Using the asymptotic expansion of $H_1^{(1)}(x)$ \cite{abramowitz1965}, the leading order approximation of $\bm \alpha$, valid far from the core ($r \gg 1$), is given by
\begin{equation}\label{eq:alpha_asymptotic}
\begin{split}
\bm \alpha (\bm r, t) &\approx \left[  \vartheta_{1y}(\varphi) - i \vartheta_{1x}(\varphi) \cos \eta \right] \\
&\times \sqrt{\frac{\pi}{2 r}} \sum_{j = 1}^{3} \frac{\bm D_j}{k_j^{3/2}} e^{i(k_j r - \omega_{BS} t - \pi / 4)}.
\end{split}
\end{equation}
This shows that far from the origin the solution indeed consists of waves propagating away from the vortex, as we claimed above.

\subsection{High-field approximation}

We shall now consider Eq.\ (\ref{eq:EquationOfMotionSource}) in the limit of high magnetic field, $\omega_L \gg 1$.
We are again interested in a solution that is periodic in time. We therefore expand $\bm \theta_2$ in Fourier series as
\begin{equation}
\bm \theta_2 \left( \bm r, t \right) = \sum_{n = -\infty}^\infty \bm \beta_n \left( \bm r  \right) e^{i n \omega_{BS} t}.
\end{equation}
Since $\bm \theta_2 \left( \bm r, t \right) $ is real, the coefficients must satisfy the relation $\bm \beta_n(\bm r) = \bm \beta_{-n}^*(\bm r)$.
Using the expression of $\bm{\hat{n}}$ from Eq.\ (\ref{eq:nt}), the coefficients $\mathsf{L}(t)$ and $\bm w(t)$ in (\ref{eq:EquationOfMotionSource}) can be written as
\begin{align}
\mathsf{L}(t) &= \sum_{n = -2}^{2} \widetilde{\mathsf{L}}_n e^{i n \omega_{BS} t}, \\
\bm w(t) &= \bm{\hat{B}} + \omega_{BS}^{-2} \sum_{n = -1}^{1} \bm{\widetilde{w}}_n e^{i n \omega_{BS} t},
\end{align}
where $\widetilde{\mathsf{L}}_n = \widetilde{\mathsf{L}}_{-n}^*$ and $\bm{\widetilde{w}}_n = \bm{\widetilde{w}}_{-n}^*$.
Plugging these into (\ref{eq:EquationOfMotionSource}) yields an infinite system of coupled partial differential equations,
\begin{equation}
\widetilde{\mathsf{K}}_n(\bm \nabla) \cdot \bm \beta_n (\bm r) + \sum_{m = -\infty}^{\infty} \widetilde{\mathsf{K}}_{n,m} \cdot \bm \beta_m (\bm r) = \bm{\widetilde{\rho}}_n (\bm r),
\end{equation}
where
\begin{align}
\widetilde{\mathsf{K}}_n(\bm \nabla) &= -(1+c) \nabla^2 + c \bm \nabla \bm \nabla - \omega_{BS}^2 \left( n^2 \mathsf{I} + i n [\bm{\hat{B}}]_\times \right), \\
\widetilde{\mathsf{K}}_{n,m} &= - i m [\bm{\widetilde{w}}_{n-m}]_\times + \widetilde{\mathsf{L}}_{n-m}, \\
\bm{\widetilde{\rho}}_n (\bm r) &= - \widetilde{\mathsf{L}}_n \cdot \bm \theta_1(\bm r).
\end{align}

In the high-field limit we may approximate $\omega_{BS} \approx \omega_L$ for all $\beta$.
Furthermore, the constant term in $\widetilde{\mathsf{K}}_n(\bm \nabla)$, proportional to $\omega_{BS}^2$, dominates all the terms $\widetilde{\mathsf{K}}_{n,m}$, except when $n = 0$. We therefore assume that the coupling terms between different $\bm \beta_n$:s may be neglected when $n \neq 0$, and are left with 
\begin{align}
\widetilde{\mathsf{K}}_0(\bm \nabla) \cdot \bm \beta_0 (\bm r) + \sum_{m = -\infty}^{\infty} \widetilde{\mathsf{K}}_{0,m} \cdot \bm \beta_m (\bm r) &= \bm{\widetilde{\rho}}_0 (\bm r), \\
\widetilde{\mathsf{K}}_n(\bm \nabla) \cdot \bm \beta_n (\bm r) &= \bm{\widetilde{\rho}}_n (\bm r), & n \neq 0
\end{align}

The time-independent part $\bm \beta_0$ will not carry energy, and so we will ignore it. Because of the symmetry $\bm \beta_n(\bm r) = \bm \beta_{-n}^*(\bm r)$, we will only consider $n < 0$. The equations are again easy to solve in the Fourier space, giving us the solutions of the form
\begin{equation}\label{eq:sol_Highfield_kspace}
\bm \beta_n (\bm k) = \widetilde{\mathsf{K}}^{-1}_n(i \bm k) \cdot \bm{\widetilde{\rho}}_n (\bm k) = \frac{1}{k} \sum_{j = 1}^3 \frac{\bm E_{n,j} \left( \varphi_k  \right)}{k^2 - k_{n,j}^2\left( \varphi_k  \right)}.
\end{equation}
Note that the poles $k_{n,j}$ now depend on the angle $\varphi_k$ and so the phase velocities of the waves depend on the direction of propagation. 

When taking the inverse Fourier transform we use a different technique than in the case of the isotropic approximation. This is due to the fact the $k_{n,j}$ depend on the angle $\varphi_k$, which makes the exact integration over $\varphi_k$ difficult. 
We shall here only calculate the asymptotic solution, valid for $r \gg 1$, since this is sufficient to calculate the energy carried by the spin waves.

The inverse Fourier transform of $\bm \beta_n(\bm k)$ is given by
\begin{equation}
\begin{split}
\bm \beta_n(\bm r) &= \frac{1}{\left( 2 \pi \right)^2} \int_0^{2 \pi} d \varphi_k \int_0^\infty k dk e^{i k r \cos(\varphi - \varphi_k)} \bm \beta_n(k, \varphi_k) \\
&=  \frac{1}{\left( 2 \pi \right)^2} \int_0^{2 \pi} d \varphi_k \int_0^\infty k dk e^{i k r \cos \varphi_k} \bm \beta_n(k, \varphi_k + \varphi),
\end{split}
\end{equation}
Here we made a change of variables $\varphi_k \to \varphi_k + \varphi$ and used the $2 \pi$-periodicity of the integrand to shift the limits of integration back to the interval $[0,2 \pi]$.
The integral over $k$ can be calculated by extending it to the complex plane. First, the poles of $\bm \beta_n$ on the real axis are shifted slightly away from the axis, $k_{n,j} \to k_{n,j} \pm i \varepsilon$, $\varepsilon > 0$. We choose the positive sign, since it produces waves propagating away from the origin when $n < 0$.

If $\cos \varphi_k > 0$, we integrate over the contour $C_+ = [0 , R] \cup C_R^+ \cup [i R, 0]$, where $C_R^+ = \{R e^{i t} | t \in [0, \pi/2]\}$ is an arc of a circle of radius $R>0$ in the first quadrant.
If $\cos \varphi_k < 0$, we use the contour $C_- = [0 , R] \cup C_R^- \cup [-i R, 0]$, where  $C_R^- = \{R e^{i t} | t \in [0, -\pi/2]\}$ is an arc of a circle of radius $R > 0$ in the fourth quadrant.
In the limit $R \to \infty$ the integral over $C_R^\pm$ vanishes due to Jordan's lemma \cite{arfken1970}. Furthermore, in the limit $R \to \infty$ and $r \to \infty$ the integral over $[\pm i R, 0]$ tends to zero sufficiently fast as a function of $r$ so that we may neglect it. Thus $\int_{C_\pm} dk \approx \int_0^\infty dk$.

On the other hand, the integral over $C_\pm$ can be calculated using the residue theorem \cite{arfken1970}. In the limit $\varepsilon \to 0$ and $r \to \infty$ the dominant contribution comes from the poles on the real axis.
Thus we obtain
\begin{equation}
\bm \beta_n \left( \bm r \right) \approx \frac{i}{4 \pi} \sum_{j} \int_{-\frac{\pi}{2}}^{\frac{\pi}{2}} d \varphi_k  e^{i k_{n,j} (\tilde{\varphi}_k) r \cos \varphi_k} \frac{\bm E_{n,j} \left( \tilde{\varphi}_k  \right)}{k_{n,j} (\tilde{\varphi}_k)},
\end{equation}
when $n < 0$.
Here $\tilde{\varphi}_k = \varphi_k + \varphi$, and the sum is calculated over those values of $j$ for which $k_{n,j}^2 > 0$.

The integral over $\varphi_k$ can be calculated using the stationary phase approximation \cite{bleistein2010} which states that when $r \gg 1$, the dominant contribution to the integral comes from the points where the derivative of the phase $\Psi_{n,j} \left( \varphi_k \right) = k_{n,j} (\varphi_k + \varphi) \cos \varphi_k$ vanishes. In our case there is only one such stationary point for each $n$ and $j$ in the interval $\left[ -\pi / 2, \pi / 2  \right]$. We denote it by $\Phi_{n,j} \left( \varphi \right)$, so that $\Psi'_{n,j} \left( \Phi_{n,j} \left( \varphi \right) \right)= 0$. Note that the stationary point varies with $\varphi$. As a final result we get
\begin{equation}
\bm \beta_n \left( \bm r \right) \approx \frac{i}{\sqrt{8 \pi r}} \sum_{j} \frac{\exp \left\{ i \left[ \Psi_{n,j} r + \frac{\pi}{4} \sgn \left( \Psi''_{n,j}  \right) \right]  \right\} }{k_{n,j} \sqrt{\left| \Psi''_{n,j}  \right|}} \bm E_{n,j},
\end{equation}
when $n < 0$. Here $\Psi_{n,j} = \Psi_{n,j} (\Phi_{n,j})$, $k_{n,j} = k_{n,j} (\varphi + \Phi_{n,j} )$, $\bm E_{n,j} = \bm E_{n,j}  (\varphi + \Phi_{n,j} )$, and $\Psi''_{n,j} = \Psi''_{n,j} (\Phi_{n,j})$ is the second derivative of the phase evaluated at the stationary point.

\section{Energy flux}\label{sec:energy}

In the preceding section we solved the equation of motion for $\bm \theta$ in two different approximations. In both cases we saw that the asymptotic solution is given by a sum of cylindrical waves propagating away from the vortex axis. In this section we calculate the amount of energy carried by these waves.
We take the unit of energy to be $\chi \Omega v R_0^2 / \mu_0 \gamma_0^2$ in our calculations. The units of any related quantities can then be easily determined from the units of energy, length, and time. For example, the unit of power per vortex length is $(\chi \Omega v R_0^2 / \mu_0 \gamma_0^2) \times (1/\Omega)^{-1} \times (v/\Omega)^{-1} = \chi \Omega^3 R_0^2 / \mu_0 \gamma_0^2$.

The amount of energy $E$ stored in the system inside a volume $V$ is given by
\begin{equation}
E = \int_V dV \mathcal{H},
\end{equation}
where 
\begin{equation}
\mathcal{H} = \bm{\dot{\theta}} \cdot \frac{\partial \mathcal{L}}{\partial \bm{\dot{\theta}}} - \mathcal{L} = \frac{\chi}{2 \mu_0 \gamma_0^2} \left( |  \bm{\dot{\theta}}|^2 - \omega_{BS}^2 | \bm w |^2  \right) + f_D + f_G
\end{equation}
is the Hamiltonian density. The rate of change of energy is then given by
\begin{equation}\label{eq:dE/dt}
\begin{split}
\frac{d E}{dt} &= \int_V dV \frac{\partial \mathcal{H}}{\partial t} = - \int_A \bm{\Sigma} \cdot d\bm A +   \int_V dV \, p,
\end{split}
\end{equation}
where $A$ is the surface of $V$, 
\begin{equation}
\Sigma_i = - (1+c) \dot{\theta}_k \partial_i \theta_k + c \dot{\theta}_k \partial_k \theta_i,
\end{equation}
and
\begin{equation}
\begin{split}
p &= \frac{2}{5} (1 - \cos \theta_0) (4 \cos \theta_0 + 3) \left( \bm{\hat{n}} \cdot \bm \theta  \right) ( \bm{\dot{\hat{n}}} \cdot \bm \theta  ) \\
&- \frac{2}{15} \sin \theta_0 (1 + 4 \cos \theta_0)  \bm{\dot{\theta}} \cdot \left( \bm{\hat{n}} \times \bm \theta  \right) \\
&- \frac{4}{15} \frac{v}{\Omega R_0} \sin \theta_0 (1 + 4 \cos \theta_0) \frac{d}{dt} \left( \bm{\hat{n}} \cdot \bm \theta  \right) .
\end{split}
\end{equation}
We see that two contributions affect the amount of energy inside $V$. The volume integral of $p$ describes the energy pumped into the system by the Brinkman-Smith mode which drives the system. The surface integral of $\bm \Sigma$ gives the energy flow through the surface of $V$. Equation (\ref{eq:dE/dt}) expresses the conservation of energy. It is analogous to Poynting's theorem in electromagnetism \cite{landau1971}, with $\bm \Sigma$ playing the part of the Poynting vector, i.e., the energy flux density vector.

When solving the equation of motion, we assumed $\bm \theta$ to be periodic in time. This means that we study the system in dynamic equilibrium. We therefore expect that the time-averaged power, $\left\langle dE / dt  \right\rangle_t$, vanishes. This is indeed so. In dynamic equilibrium the energy absorbed into the system inside the volume $V$ is equal to the energy flux through the surface of $V$.

Because the vortex is uniform in the $z$-direction, we choose the volume $V$ to be a cylinder of radius $r$ with its axis on the vortex axis. Let us denote the amount of energy absorbed into the system inside the cylinder per unit time and vortex length, averaged over time, by $P_{a}(r)$. Similarly, let us denote the time-averaged energy flux per vortex length out of the cylinder by $P_{f}(r)$. Based on the above discussion, these are both equal. We call this common value $P(r)$, so that
\begin{equation}
P(r) = P_{a}(r) = P_{f}(r) = \int_0^{2 \pi} d \varphi \sigma_r \left( r, \varphi \right) ,
\label{e.sspow}\end{equation}
where we have defined $\sigma_r \left( r, \varphi \right) = r \Sigma_r \left( r, \varphi \right)$.
There is no net flow of energy through the upper and lower surfaces of the cylinder because of the uniformity of the vortex along its axis.

In the following we discuss the behaviour of $\sigma_r(r, \varphi)$ and $P(r)$ as a function of different parameters. In the numerical calculations we use the coefficients $C_1 = C_2 = 1.33$ for the A-phase-core vortex and $C_1 = 3.00$, $C_2 = 0.08$ for the double-core vortex. These are the values obtained from numerical calculations at $T = 0.6 T_c$, $p = 29.3$ bar, as we mentioned in Sec.\ \ref{sec:static_vortex}. In the high-field approximation we set $c = 1$. Finally, if not stated otherwise, we use parameters $\beta = \theta_L$, $\eta = 0$, $\zeta = 0$ and $\omega_L = 2$.

\begin{figure}[tb]
\centering
\subfloat[]{\includegraphics[width=0.9\columnwidth]{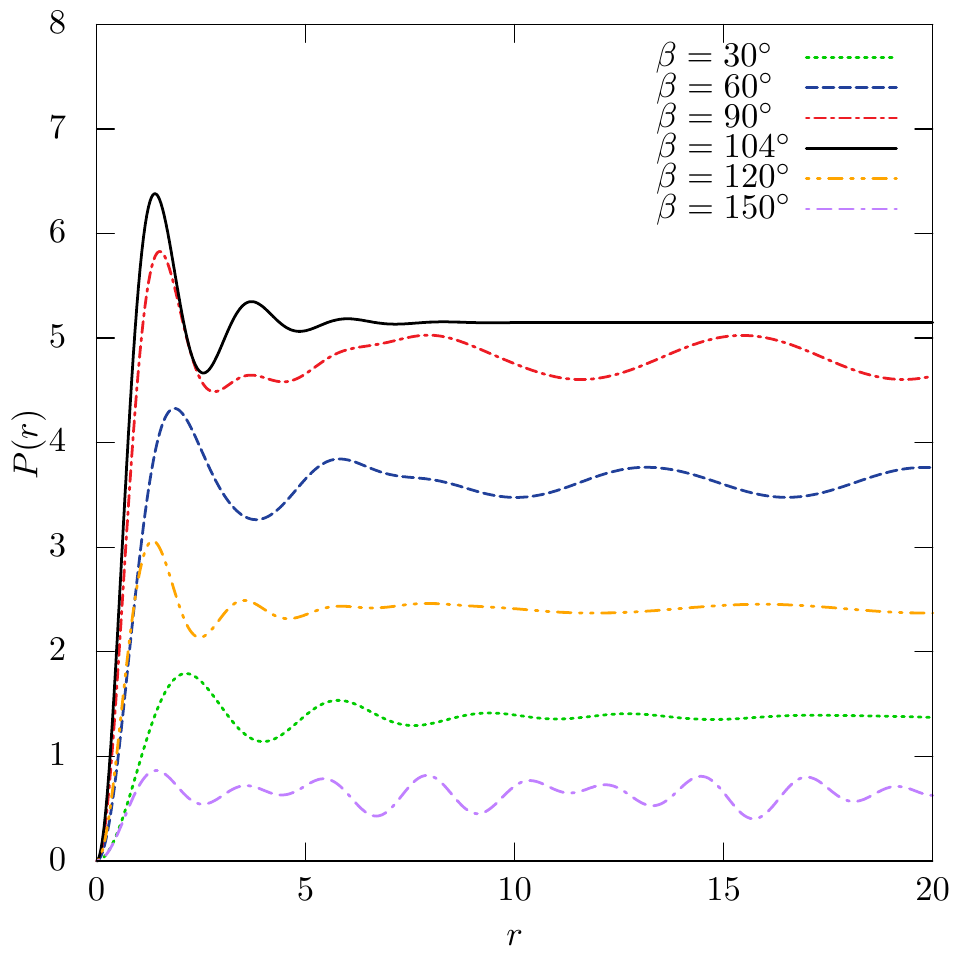}}

\subfloat[]{\includegraphics[width=0.9\columnwidth]{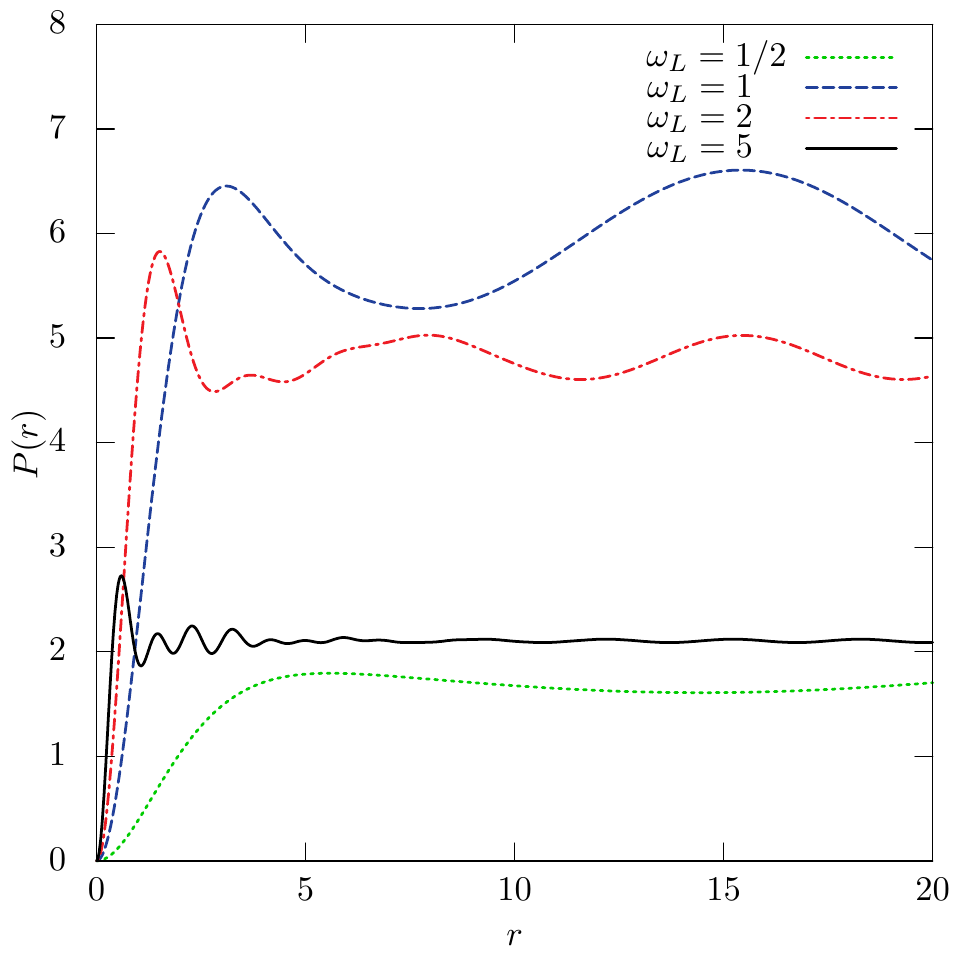}}
\caption{The steady-state power (\ref{e.sspow}) per unit length inside a cylinder of radius $r$ centered at the vortex as a function of $r$. The results shown are obtained in the isotropic approximation for the double-core vortex. (a) Fixed $\omega_L = 2$, six different values of the tipping angle $\beta$. (b) Fixed $\beta = 90^\circ$, four different values of $\omega_L$. The values of $C_1$, $C_2$, $\zeta$, and $\eta$ are given in Sec.\ \ref{sec:energy}.}
\label{fig:Pr}
\end{figure}

Let us start by considering the dependence of $\sigma_r(r, \varphi)$ and $P(r)$ on $r$. 
Figure \ref{fig:Pr} shows $P(r)$ as a function of $r$ in the case of the double-core vortex for some values of $\omega_L$ and $\beta$ in the isotropic approximation, where we were able to solve the equation of motion for all $r$.
The exact form of $P(r)$ depends on the parameters used, but the general trend is clear.
$P(r)$ starts from zero at the origin and increases monotonically within the range of a few dipole lengths $\xi_D = \sqrt{15} / 2$. Then, within the next few dipole lengths, there are transient oscillations. Finally, when $r$ is large, $P(r)$ oscillates about some average value. These asymptotic oscillations stem from interference between different wave modes in Eq.\ (\ref{eq:alpha_asymptotic}). Their amplitude depends on $\omega_L$ and $\beta$, and is at its largest somewhere near $\beta = 90^\circ$, $\omega_L = 1$. There are no asymptotic oscillations when $\beta = \theta_L$ since there is only one wave mode present. The oscillation amplitude approaches zero at large $\omega_L$. This is in accordance with the high-field approximation, which predicts that $P(r)$ is independent of $r$. The behaviour of the A-phase-core vortex is qualitatively similar.

From the form of $P(r)$ we see that most of the energy is absorbed into the system from the region of radius $\sim \xi_D$ around the vortex core.
This is smaller than the usual inter-vortex distance in the experiments, which is $\sim 10 \xi_D$. Combining this to the fact that the asymptotic oscillations of $P(r)$ are, at least in most cases, relatively small, we can focus our interest on the average value of $P(r)$ at large $r$. We denote this average value by
\begin{equation}
P \equiv \lim_{r \to \infty} \left\langle P(r) \right\rangle_r.
\end{equation}
Similarly, we denote the average value of $\sigma_r(r,\varphi)$ by
\begin{equation}
\sigma_r(\varphi) \equiv \lim_{r \to \infty} \left\langle \sigma_r(r,\varphi) \right\rangle_r.
\end{equation}
These are related by
\begin{equation}
P = \int_0^{2 \pi} d \varphi \sigma_r \left(\varphi \right).
\end{equation}

\begin{figure}[tb]
\centering
\subfloat[]{\includegraphics[width=0.9\columnwidth]{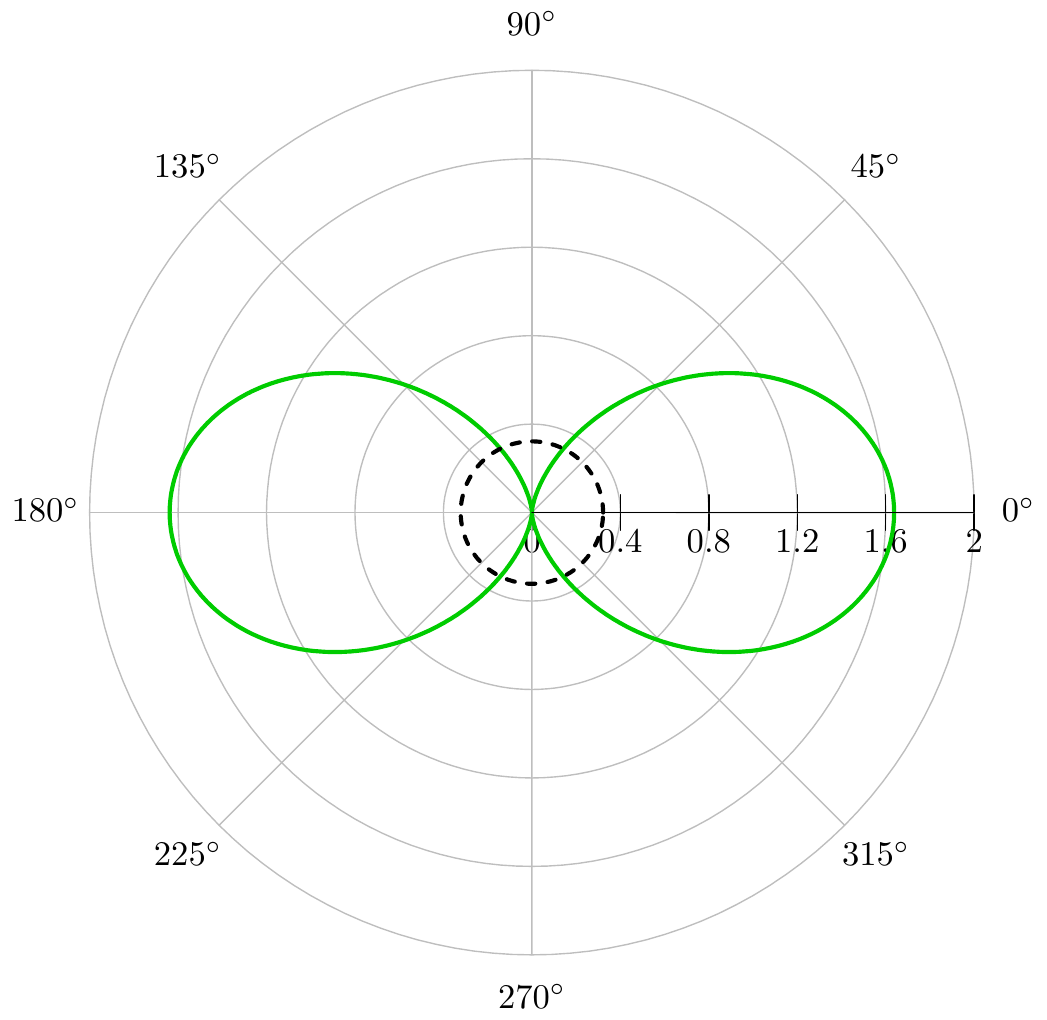}}

\subfloat[]{\includegraphics[width=0.9\columnwidth]{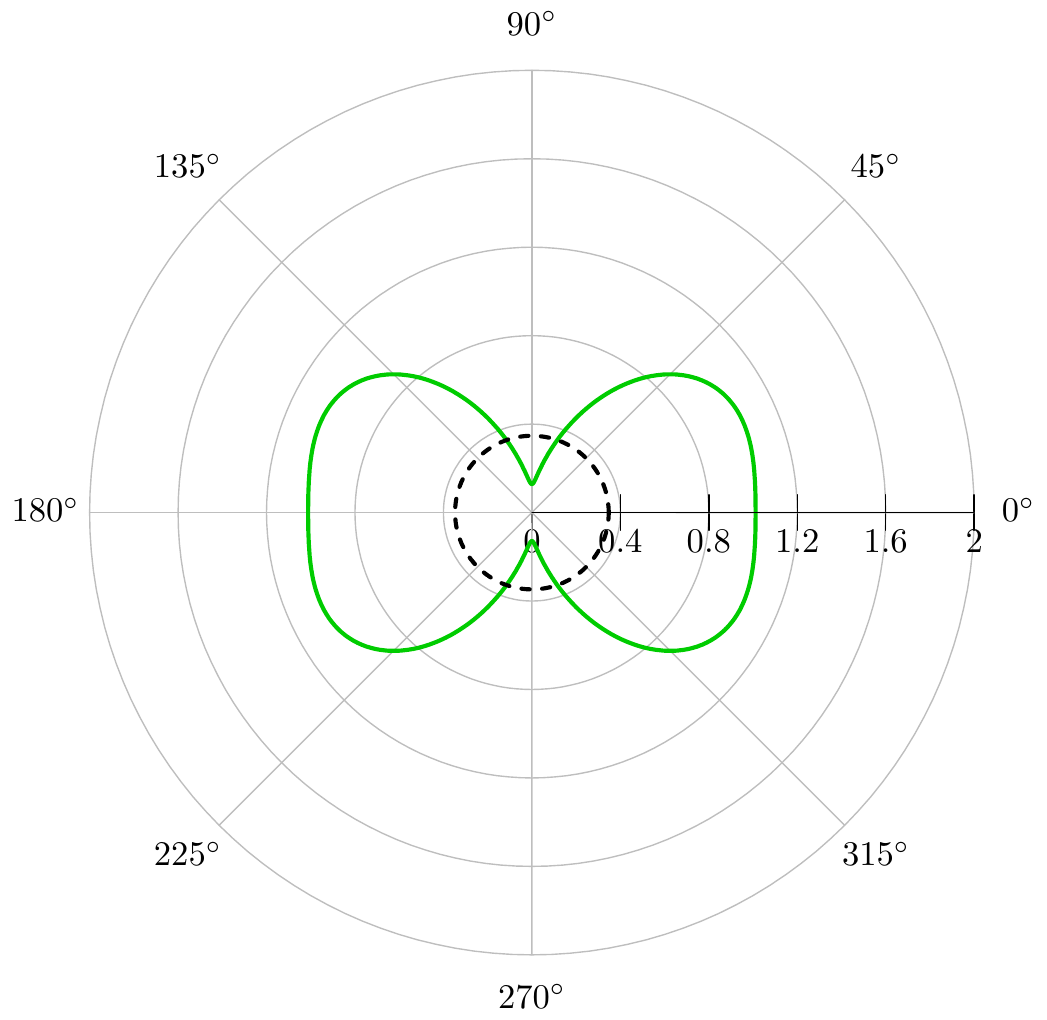}}
\caption{Radiation pattern of vortices, i.e., the angular dependence of $\sigma_r \left( \varphi \right)$. Distance from the origin at a given angle represents the amount of energy flowing in that direction. The dashed black line represents the A-phase-core vortex, while the solid green line represents the double-core vortex. (a) Isotropic approximation. (b) High-field approximation. The values of $C_1$, $C_2$, $c$, $\beta$, $\zeta$, $\eta$, and $\omega_L$ are given in Sec.\ \ref{sec:energy}.}
\label{fig:Sigmar}
\end{figure}

The explicit form of $P$ is, in general, inconveniently complicated. One exception is the case $\beta = \theta_L$, $\eta = 0$. In this case we have
\begin{equation}\label{eq:P_c0}
P =\frac{\pi^2}{8} \omega_L \frac{2 \omega_L^2 - 1 + \sqrt{1 + 4 \omega_L^4}}{1 + 4 \omega_L^4} \left( C_1^2 + C_2^2  \right)
\end{equation}
in the isotropic approximation and
\begin{equation}\label{eq:P_HighField}
\begin{split}
P = \frac{\pi^2}{8} \omega_L^{-1} &\Bigg[ \frac{3 c^2 + 6 c + 4 }{4 \left(1 + c \right)^2} \left( C_1^2 + C_2^2  \right) + \frac{2 c \left( 2 + c \right)}{4 \left(1 + c \right)^2} C_1 C_2 \Bigg]
\end{split}
\end{equation}
in the high-field approximation. Note that the high-field limit of Eq.\ (\ref{eq:P_c0}) coincides with Eq.\ (\ref{eq:P_HighField}) when $c = 0$, as it should. Figure \ref{fig:Sigmar} shows the radiation pattern, i.e., the angular dependence of $\sigma_r \left( \varphi \right)$. The A-phase-core vortex radiates symmetrically in both approximations. The pattern of the double-core vortex, on the other hand, is highly anisotropic. Most of the energy flow is in the direction perpendicular to $\bm{\hat{b}}$, with only a small fraction of the flow in the direction of $\bm{\hat{b}}$. We also see that the shape of the pattern is different in the two approximations. This stems from the different values of $c$ used in the approximations.

\begin{figure}[tb]
\centering
\subfloat[]{\includegraphics[width=0.9\columnwidth]{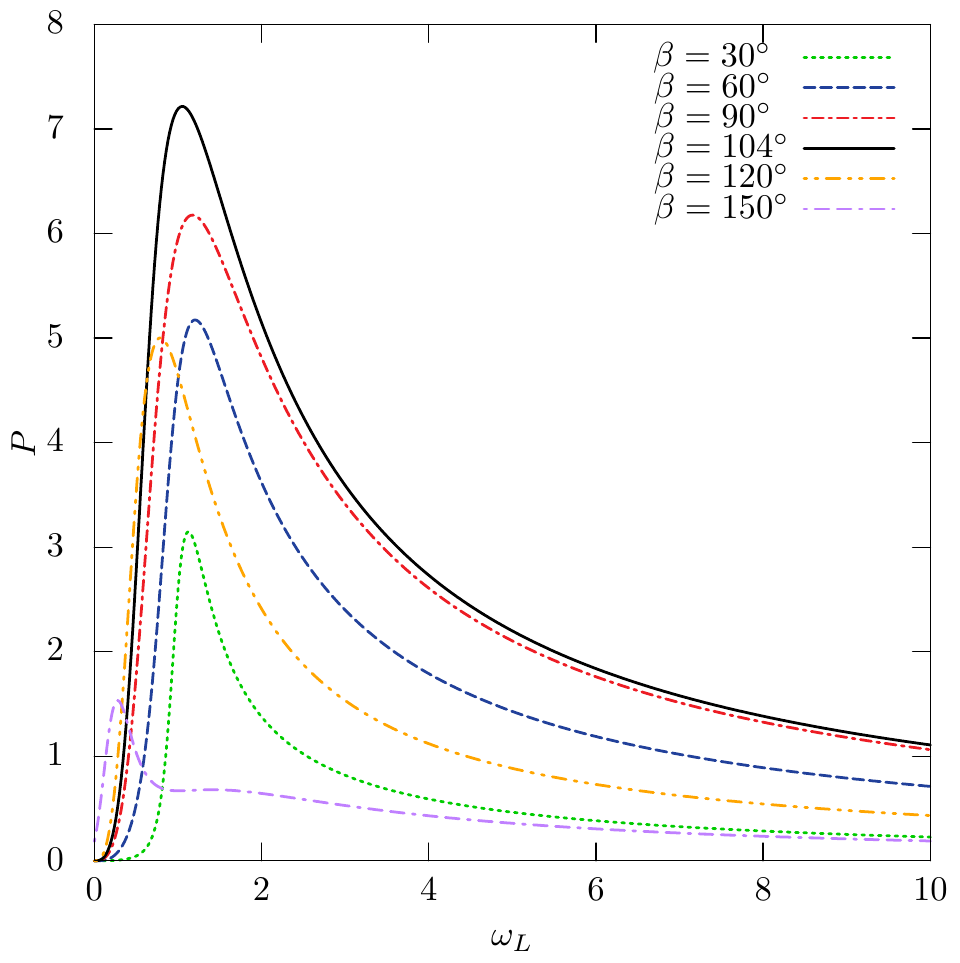}}

\subfloat[]{\includegraphics[width=0.9\columnwidth]{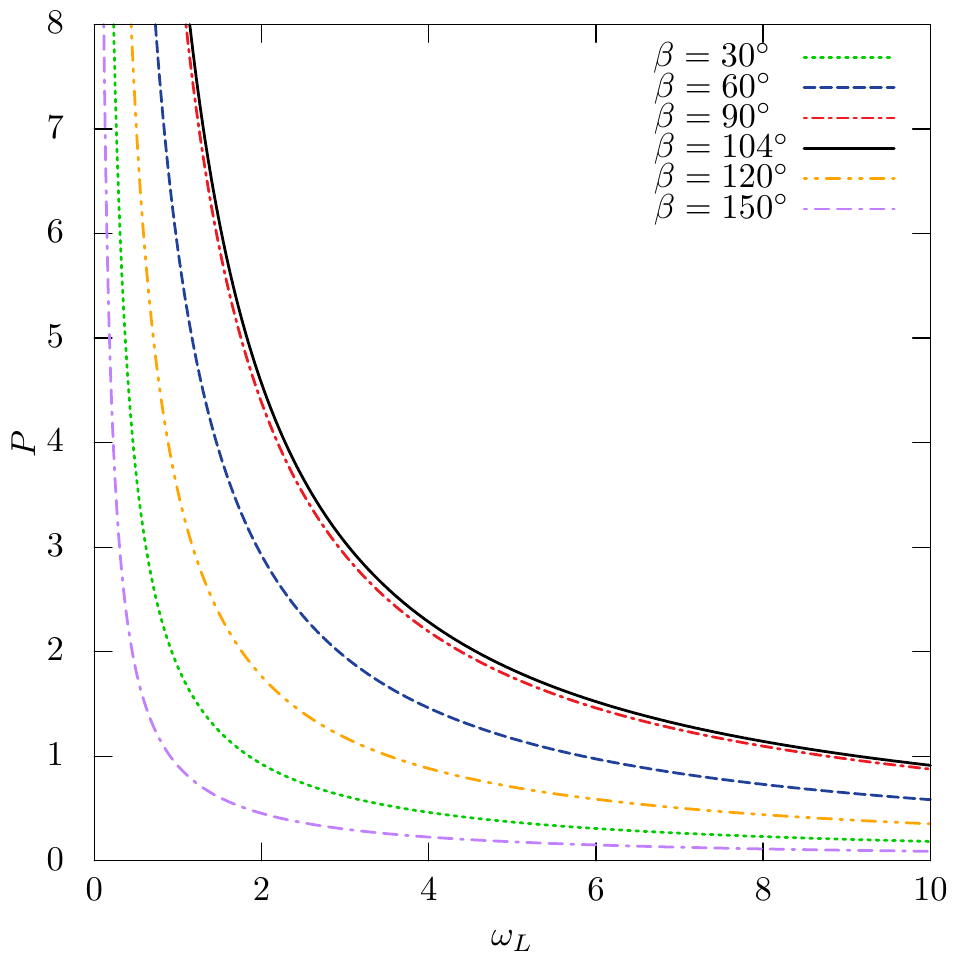}}
\caption{Radiated power per vortex length as a function of $\omega_L$ for the double-core vortex at six different values of the tipping angle $\beta$. (a) Isotropic approximation. (b) High-field approximation. The values of $C_1$, $C_2$, $c$, $\zeta$, and $\eta$ are given in Sec.\ \ref{sec:energy}.}
\label{fig:PTippedMagnetisation}
\end{figure}

Figure \ref{fig:PTippedMagnetisation} shows the behaviour of $P$ as a function of $\omega_L$ for different tipping angles $\beta$ in the case of the double-core vortex. In the high-field approximation $P \propto \omega_L^{-1}$. In the isotropic approximation $P$ behaves similarly for large $\omega_L$, but the low-field behaviour is different. The power vanishes at $\omega_L = 0$ and has a maximum near $\omega_L = 1$. In both approximations $P$ is an increasing function of $\beta$ up to $\theta_L$, beyond which it starts to decrease. The behaviour of the A-phase-core vortex is qualitatively similar.

\begin{figure}[tb]
\centering
\subfloat[]{\includegraphics[width=0.9\columnwidth]{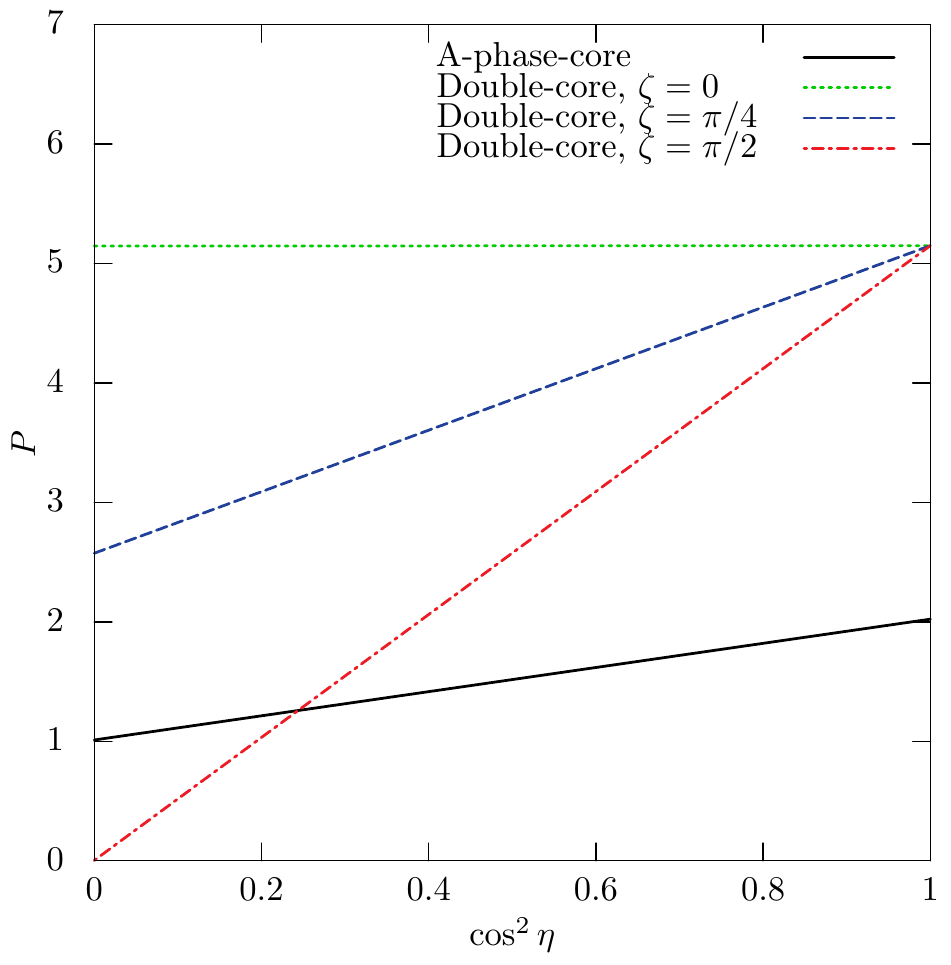}}

\subfloat[]{\includegraphics[width=0.9\columnwidth]{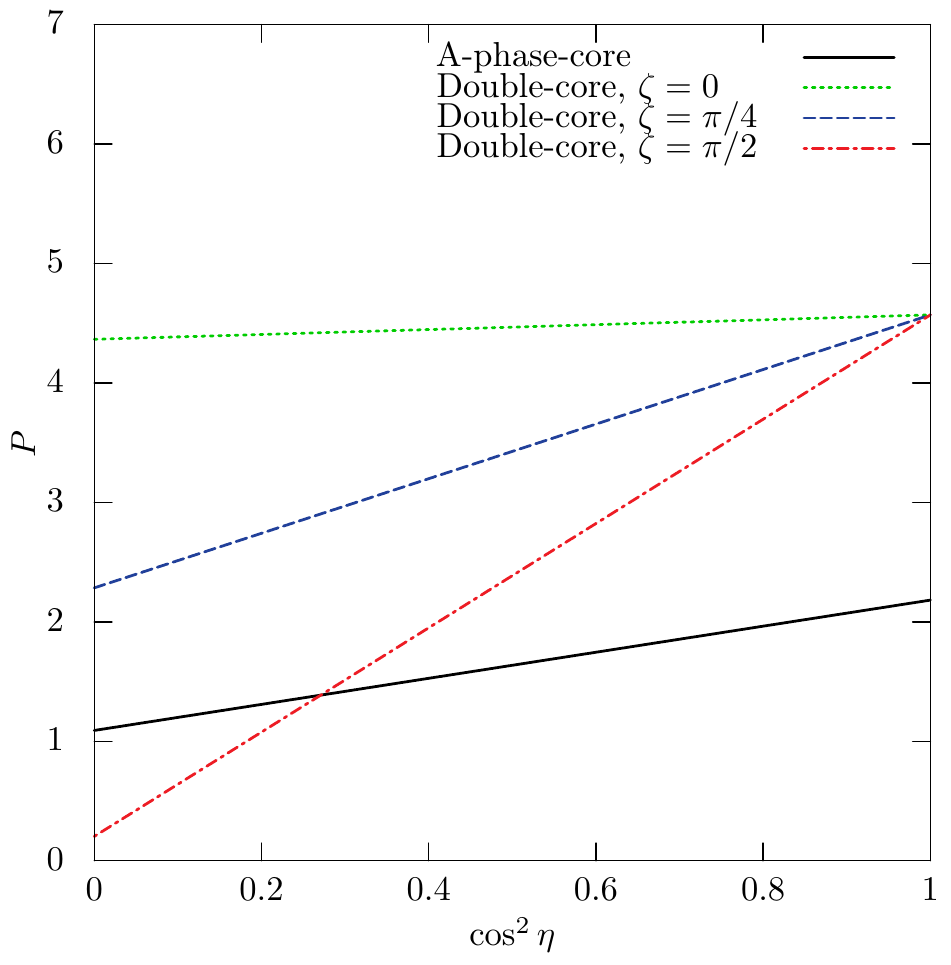}}
\caption{Radiated power per vortex length as a function of $\cos^2 \eta$ at three different values of $\zeta$. The result for the A-phase-core vortex is independent of $\zeta$ due to symmetry of the vortex. (a) Isotropic approximation. (b) High-field approximation. The values of $C_1$, $C_2$, $c$, $\beta$, and $\omega_L$ are given in Sec.\ \ref{sec:energy}.}
\label{fig:PTiltedField}
\end{figure}

Another interesting case to study is the dependence of $P$ on the direction of the magnetic field. Figure \ref{fig:PTiltedField} shows $P$ as a function of $\cos^2 \eta$ at three different values of $\zeta$. As noted before, the A-phase-core vortex is symmetric and thus $P$ is independent of $\zeta$. The result for the double-core vortex, on the other hand, is highly dependent on $\zeta$. The susceptibility anisotropy of the double-core vortex favours the orientation $\zeta = \pi / 2$ in tilted field \cite{thuneberg1987}. Note that in all cases $P\left( \eta  \right) = a_0 + a_2 \cos^2 \eta$ with some constants $a_0$ and $a_2$. 

\section{Twisted vortex}\label{sec:twisted}

As mentioned in Sec.\ \ref{sec:introduction}, it is possible that the precessing magnetization of the Brinkman-Smith mode can rotate the half cores of the double-core vortex around each other, causing the vortex to twist. 
In this section we study how the radiation of spin waves is affected by uniform twisting of the core. This can be modelled by assuming that $\zeta$ depends on $z$ as $\zeta (z) =  \kappa z$, where $\kappa$ is a dimensionless constant describing the amount of twisting. As a result, $\bm \theta_1$ will also depend on $z$ and there will be a new term in the equation of motion from the derivatives of $\bm \theta_1$ with respect to $z$, see Eqs.\ (\ref{eq:rho}) and (\ref{eq:theta_1}). For simplicity, we shall discuss here only the case $\beta = \theta_L$, $\eta = 0$. 

\begin{figure}[tb]
\centering
\subfloat[]{\includegraphics[width=0.49\columnwidth]{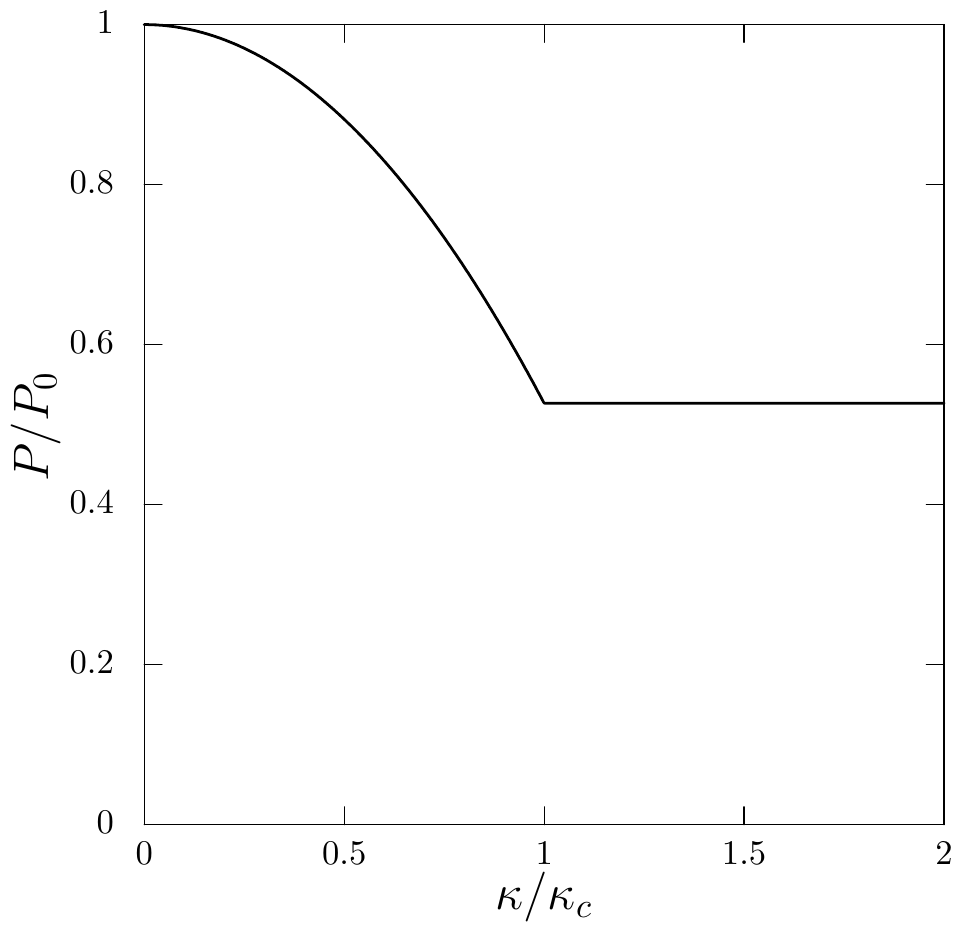} \label{fig:PTwisting}}
\subfloat[]{\includegraphics[width=0.49\columnwidth]{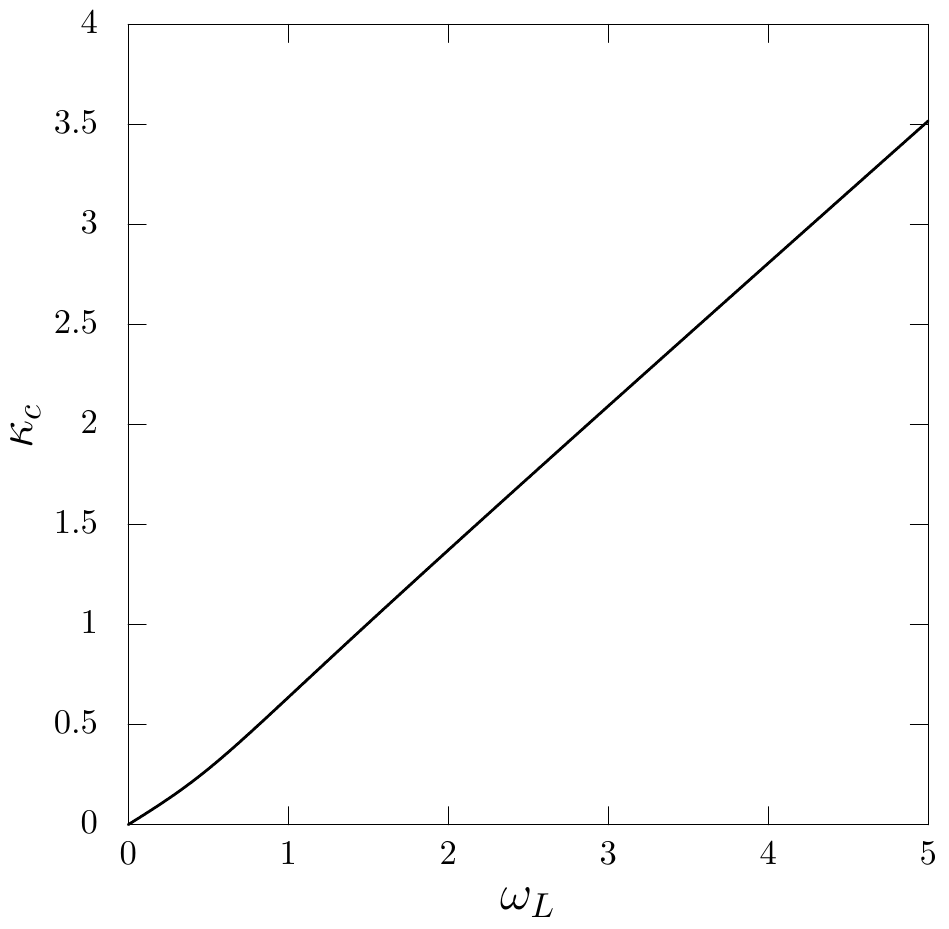} \label{fig:Kappac}}
\caption{(a) The dependence of the radiated power on twisting of the double-core vortex in the isotropic approximation, Eq.\ (\ref{eq:P_Twisted}). (b) The dependence of $\kappa_c$ on $\omega_L$ in the isotropic approximation, Eq.\ (\ref{eq:kappa_c}).}
\label{fig:Twisting}
\end{figure}

In the isotropic approximation, when there is no twisting, there is only one wave mode present, with wavenumber $k_0$. 
When twisting increases, the solution is of the form $\bm \alpha \left( \bm r, t  \right) = e^{-i \omega_{BS} t}\left[ \bm \beta_1\left( r, \varphi  \right) + e^{-2 i \kappa z}  \bm \beta_2\left( r, \varphi  \right) \right]$. Here $ \bm \beta_1$ describes a wave with the original wave number $k_0$, while $ \bm \beta_2$ describes a wave with a wavenumber $k = \sqrt{ k_0^2 - 4 \kappa^2}$. Thus, when the twisting increases, there is a critical value $\kappa_c = k_0 / 2$ beyond which $k$ becomes imaginary. Since $\bm \beta_1$ is not affected by twisting, it is the only part of the solution that carries energy away from the vortex when $\kappa > \kappa_c$. 
The power per vortex length is given by
\begin{equation}\label{eq:P_Twisted}
\begin{split}
P / P_0 =
\begin{cases}
1 - \frac{1}{2} \frac{\kappa^2}{\kappa^2_{c}} \frac{ \left( C_1 - C_2  \right)^2}{C_1^2 + C_2^2}, & 0 \leq \kappa \leq \kappa_c \\
\frac{1}{2}\frac{(C_1 + C_2)^2}{C_1^2 + C_2^2}, & \kappa > \kappa_c
\end{cases},
\end{split}
\end{equation}
where
\begin{equation}
P_0 = \frac{\pi^2}{8} \omega_L \frac{2 \omega_L^2 - 1 + \sqrt{1 + 4 \omega_L^4}}{1 + 4 \omega_L^4}  \left( C_1^2 + C_2^2  \right)
\end{equation}
is the value of $P$ for an untwisted vortex and
\begin{equation}\label{eq:kappa_c}
\kappa_c = \sqrt{ \frac{2 \omega_L^2 - 1 + \sqrt{1 + 4 \omega_L^4}}{8} }.
\end{equation}
First of all we see that $P$ is independent of $\kappa$ in the case of the A-phase-core vortex $\left( C_1 = C_2 \right)$ so only the double-core vortex is affected by twisting. This is again due to the cylindrical symmetry of the A-phase-core vortex.
We also see that the result coincides with our earlier result (\ref{eq:P_c0}) when $\kappa = 0$. When $\kappa \leq \kappa_c$, power decreases quadratically with $\kappa$. When $\kappa > \kappa_c$, $P$ is constant. Figure \ref{fig:Twisting}(a) shows the ratio $P / P_0$ as a function of $\kappa / \kappa_c$ at $\omega_L = 2$. Figure \ref{fig:Twisting}(b) shows $\kappa_c$ as a function of $\omega_L$.

In the high-field approximation there are more wave modes present when $c \neq 0$. This makes things more complicated. It is, however, easy to calculate what is the maximal effect of twisting. Two of the modes are independent of $\kappa$. The remaining ones all have a critical value $\kappa_c^{(i)}$, so that the $i$:th mode disappears when $\kappa > \kappa_c^{(i)}$. When $\kappa > \kappa_c \equiv \max\{  \kappa_c^{(i)} \}$, the power attains its minimum value
\begin{equation}
P_{\min} = \frac{\pi^2}{16} \omega_L^{-1} \left(  C_1 + C_2  \right)^2.
\end{equation} 
This is in accordance with the high-field limit of Eq.\ (\ref{eq:P_Twisted}).

Both approximations therefore show the same qualitative behaviour. Twisting of the vortex core reduces the radiated power up to some saturation point $\kappa_c$. Further twisting has no effect on the power.

\section{Comparison with experiments}\label{sec:experiments}

In this section we compare the results above with experimental results from Refs.\ \cite{kondo1991} and \cite{dmitriev1990}. We include only the dissipation by spin wave radiation in the quantitative comparisons, although we know that the Leggett-Takagi relaxation also contributes \cite{laine2016}. For simplicity, we use the isotropic approximation. For the double-core vortex we use $C_2 / C_1 = 0$, as vortex-structure calculations indicate that $C_2$ is small, and $\zeta = \pi / 2$, which is favored by susceptibility anisotropy. This leaves $C_1$ as the only free parameter. To compare theory with experiments, we first determine $C_1$ that gives the best fit to the measured values. After that, we compare the fitted value of $C_1$ with the one obtained from numerical solution of the vortex structure. Unless otherwise mentioned, the experiments were done using $p = 29.3$ bar, $B = 14.2$ mT, $\eta = 0$, and $\beta = \theta_L$.

\begin{figure}[tb]
\centering
\includegraphics[width=0.9\columnwidth]{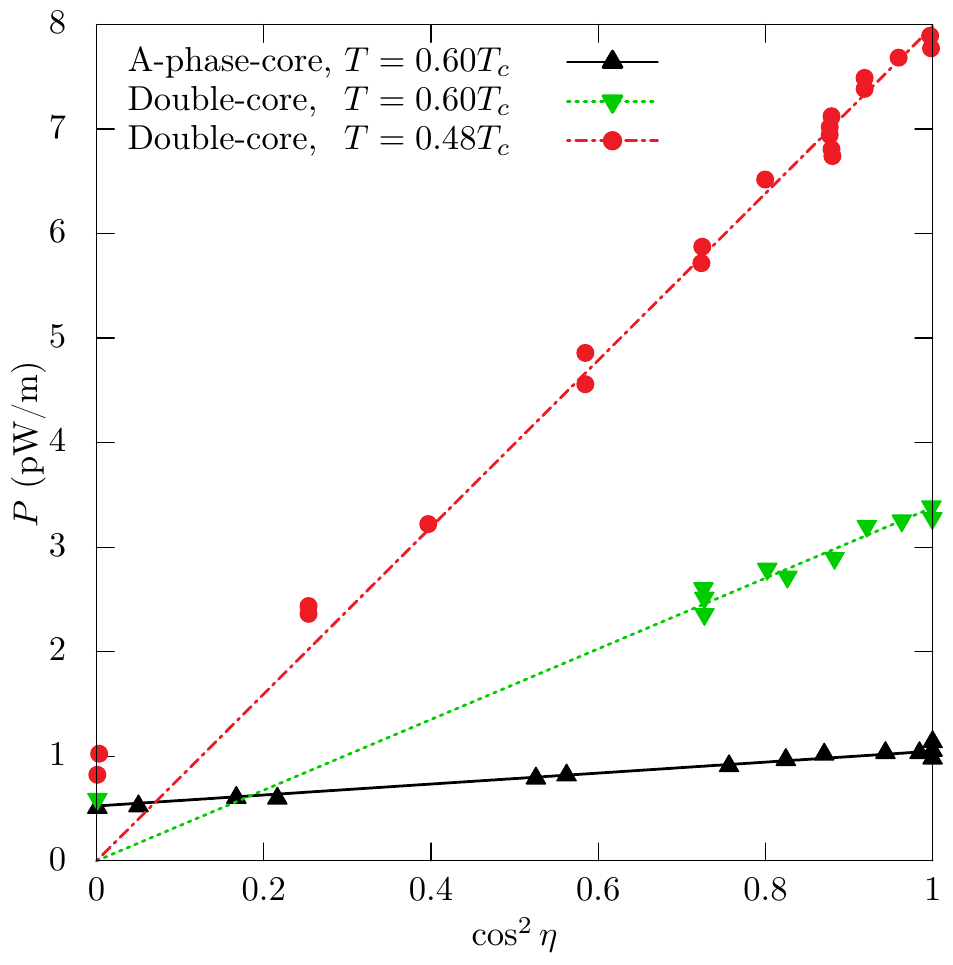}
\caption{A comparison between theoretical (lines) and experimental (points) values of $P$ as a function of $\cos^2 \eta$. The experimental data is from Ref.\ \cite{kondo1991}. Theoretical curves use $C_1$ fitted to the data. The fitting procedure yields $C_1 = 1.66$ for the A-phase-core vortex at $T = 0.60 T_c$, $C_1 = 5.81$ for the double-core vortex at $T = 0.48 T_c$, and $C_1 = 4.20$ for the double-core vortex at $T = 0.60 T_c$. Theoretical curves are of the form $P\left( \eta  \right) = a_0 + a_2 \cos^2 \eta$. The experimental data seems to obey the same formula. Since we have assumed that $C_2 = 0$ for the double-core vortex, theory predicts that $a_0 = 0$. In the experiment $a_0 > 0$. External parameters are given in Sec.\ \ref{sec:experiments}.}
\label{fig:KondoComparison}
\end{figure}

Figure \ref{fig:KondoComparison} shows the absorption per vortex length as a function of $\cos^2 \eta$. The experimental data is taken from Fig.\ 2 of Ref.\ \cite{kondo1991}. There are three different data sets shown in the figure, one for the A-phase-core vortex at $T = 0.60 T_c$ and two for the double-core vortex at temperatures $0.48 T_c$ and $0.60 T_c$. Each of these would seem to obey the rule $P\left( \eta  \right) = a_0 + a_2 \cos^2 \eta$, as noted in \cite{kondo1991}. This is also predicted by theory. Theoretical curves shown in the figure use parameter $C_1$ fitted to the experimental data. For the A-phase-core vortex we obtain $C_1 = 1.66$. For the double-core vortex we obtain $C_1 = 5.81$ at $T = 0.48 T_c$ and $C_1 = 4.20$ at $T = 0.60 T_c$. Note that since we have assumed $C_2 = 0$ for the double-core vortex, theory predicts that the constant $a_0$ vanishes, which seems to be contrary to the experimental data. Similar problem appears with the Leggett-Takagi relaxation, which also has quadratic dependence on $C_1$ and $C_2$ \cite{laine2016}.

Figure 1 in Ref.\ \cite{dmitriev1990} shows that the measured absorption decreases when the magnetic field is increased from $14.2$ mT to $28.4$ mT. We study the ratio $\varrho = P\left( 28.4 \text{mT}  \right)/P\left( 14.2 \text{mT}  \right) $ near $T_V$, which is the phase transition temperature between the core structures. The measured values are $\varrho= 0.68$ for both vortex types. Theory predicts $\varrho = 0.54$. As a comparison, the absorption by the Leggett-Takagi relaxation is field-independent, while the absorption by spin diffusion increases quadratically with the field.

\begin{figure}[tb]
\centering
\includegraphics[width=0.9\columnwidth]{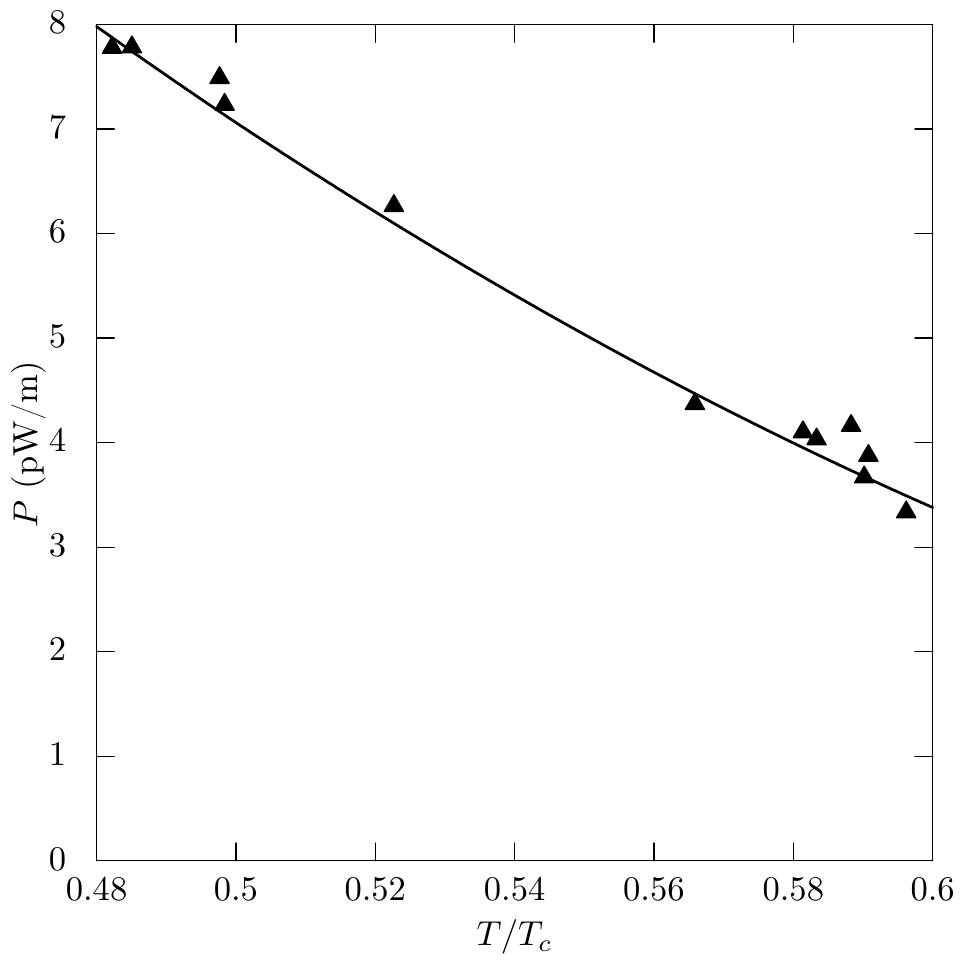}
\caption{A comparison between theoretical (line) and experimental (points) values of $P$ as a function of temperature. The experimental data is from Ref.\ \cite{kondo1991}. The theoretical curve is obtained assuming linear dependence of $C_1$ on $T$, $C_1(T) = A T / T_c + B$. The coefficients $A = -13.4$ and $B = 12.2$ were calculated using the values $C_1(0.48 T_c) = 5.81$ and $C_1(0.60 T_c) = 4.20$ that we obtained from fitting to the data as a function of $\cos^2 \eta$, see Fig.\ \ref{fig:KondoComparison}. External parameters are given in Sec.\ \ref{sec:experiments}.}
\label{fig:KondoComparisonTemperature}
\end{figure}

Figure \ref{fig:KondoComparisonTemperature} shows the absorption as a function of temperature for the double-core vortex. The experimental data is taken from Fig.\ 1 of Ref.\ \cite{kondo1991}. Since the temperature range is quite narrow, $0.48 T_c < T  < 0.6 T_c$, we assume that we can approximate $C_1$ by a linear function $C_1(T) = A T / T_c + B$. The coefficients $A$ and $B$ can be calculated using the values $C_1(0.48 T_c) = 5.81$ and $C_1(0.60 T_c) = 4.20$ we obtained above. The theoretical curve shown in the figure uses this linear approximation for $C_1$. The clear temperature dependence is in contrast to the Leggett-Takagi relaxation, where the absorption is essentially temperature-independent \cite{laine2016}.

Next we consider the effect of twisting the double-core vortex. According to Ref.\ \cite{kondo1991}, the measured ratio $\varrho_{\text{twist}}$ between the absorptions in the twisted and untwisted states of the double-core vortex is $\varrho_{\text{twist}} = 0.83$ at $T = 0.5 T_c$ and $\varrho_{\text{twist}} = 0.87$ at $T = 0.6 T_c$. These are in keeping with theory, which predicts that $1/2 \leq \varrho_{\text{twist}} \leq 1$. The measured values of $\varrho_{\text{twist}}$ correspond to $\kappa / \kappa_c = 0.58$ and $\kappa / \kappa_c = 0.51$, or $\kappa = 0.68 $ and $\kappa = 0.66 $, respectively.

Based on the numerical solution of the double-core vortex structure, the value $C_1=3.7$ was obtained in Ref.\ \cite{silaev2015} at $T=0.5 T_c$. This is by a factor of $2 / 3$ smaller than the value obtained from the fitting procedure above. Since the absorption is quadratic in $C_1$, approximately one half of the measured absorption is explained by this value of $C_1$. Better agreement is obtained in the temperature dependence. Based on the vortex structure calculation, $C_1(0.6T_c)/C_1(0.5T_c) = 0.81$. This is close to the value $0.75$ obtained above (Fig.\ \ref{fig:KondoComparisonTemperature}). 

To summarize, we have compared the theoretical model of dissipation by spin wave radiation with experiments reported in Refs.\ \cite{kondo1991} and \cite{dmitriev1990}. Without any adjustable parameters, it explains the order of magnitude of the absorption. What is more, it accounts well for the dependencies of the absorption on the direction and the magnitude of the magnetic field, on temperature, and on twisting.
With the Leggett-Takagi relaxation, only the dependence on the field direction can be understood. Thus it seems that major part of the absorption is explained by radiation of spin waves. Including both the spin wave radiation and the Leggett-Takagi relaxation in the analysis would lead to better agreement with experiments, especially in the magnitude of the absorption. Possible reasons for the remaining problems may be the inadequacy of the weak-coupling theory to calculate the parameters $C_1$ and $C_2$, as well as the omission of spin diffusion and the detailed structure of the vortex core. One task that still remains to be done is detailed comparison of twisting and its dynamics with experiments \cite{Sonin93,Krusius93}. 

Both the present model of spin wave radiation and the Leggett-Takagi relaxation calculated in 
Ref.\ \cite{laine2016} give absorption that is quadratic in the coefficient $C_1$ and $C_2$. These coefficients are solely determined by the structure of the vortex core. For example, the simplest theoretical vortex structure has $C_1=C_2=0$ \cite{Ohmi83}. This explains the experimental observation that the absorption at large tipping angles is more sensitive to the vortex-core structure than the frequency shift at small tipping \cite{Ikkala82,Hakonen89}. The latter is determined by susceptibility anisotropy, which is only partially dependent on the core structure.

\section{Conclusions}

We have studied spin dynamics of superfluid $^3$He-B in the presence of an isolated vortex.
The vortex perturbs the uniformly precessing magnetization and gives rise to spin waves. These waves carry energy, causing dissipation in the system. We calculated the amount of dissipation and its dependence on several parameters. Good agreement with experiments indicates that spin wave radiation is the dominant dissipation mechanism for vortices in the intermediate-temperature range.

\begin{acknowledgments}
We thank Volodya Eltsov, Matti Krusius, Edouard Sonin, and Grigory Volovik for discussions.
This work was financially supported by the Vilho, Yrj\"o and Kalle V\"ais\"al\"a Foundation, the Jenny and Antti Wihuri Foundation and the Oskar \"Oflunds Stiftelse sr. 
\end{acknowledgments}

\end{document}